\documentclass[12pt]{article}
\usepackage[utf8]{inputenc}
\usepackage[margin=0.8in]{geometry}
\usepackage{blindtext, xfrac}
\usepackage[english]{babel}
\usepackage[T1]{fontenc}
\usepackage{titling}
\usepackage{amsmath}
\usepackage{amssymb}
\usepackage{cite}
\usepackage{graphicx}
\usepackage{subcaption}
\usepackage{booktabs}
\usepackage{physics}
\usepackage{nameref}
\usepackage{ bbold }
\usepackage{textgreek}
\usepackage{hyperref}
\usepackage{gensymb}
\usepackage{float}
\usepackage[usenames,dvipsnames]{color}
\usepackage{svg}
\usepackage{placeins}
\usepackage{tabularx}
\definecolor{greeny}{rgb}{0.2,0.8,0.5}
\definecolor{darkyellow}{RGB}{218,145,0}
\usepackage{wrapfig}

\begin{document}
\title{New Windows on Heavy Dark Matter: Mineral Melt Modelling and X-Ray Readout for Muscovite Mica}

\author{Yilda Boukhtouchen$^{1,2}$, Joseph Bramante$^{1,2}$, Andrew Buchanan$^{1,2}$, Alexander Hayes$^{1,2}$,\\ Matthew Leybourne$^{1,3}$,
Jennika McIntosh$^{1,2}$, Anupam Ray$^{1,2,5}$, Aaron Shugar$^{4}$
\\
$^1$\small Arthur B. McDonald Canadian Astroparticle Physics Research Institute, 64 Bader Lane, \\ \small Queen's University, Kingston, Ontario, Canada
\\
$^2$\small Department of Physics, Engineering Physics, and Astronomy, \\ \small Queen's University, Kingston, Ontario, Canada
\\
$^3$\small Department of Geological Sciences and Geological Engineering,\\ \small Queen's University, Kingston, Ontario, Canada
\\
$^4$\small Department of Art History and Art Conservation, \\ \small Queen's University, Kingston, Ontario, Canada
\\
$^5$ \small Perimeter Institute for Theoretical Physics, Waterloo, ON N2J 2W9, Canada}

\maketitle

\begin{abstract}
Muscovite mica is a translucent, layered silicate mineral whose basal cleavage, low radiogenic background, gigayear exposures, and demonstrated track retention over geological timescales make it a compelling target for rare particle searches. In this work, we develop a new framework for detecting heavy composite dark matter using muscovite mica as a paleodetector. We model melt track formation by heavy composite dark matter transiting through mica using a Sedov-Taylor thermal spike formalism, and validate the sub-micron regime with SRIM/TRIM simulations of nuclear recoil cascades, which also calibrate the phonon efficiency governing local energy deposition. We demonstrate a novel readout method using rapid X-ray fluorescence mapping with a copper backing contrast technique, capable of identifying micron-scale damage features in cleaved mica sheets over macroscopic scan areas, and calibrate the minimum detectable track size using laser-ablated defect regions. We present projected sensitivities for opaque and diffuse composite dark matter, including a sub-melt hole-channel detection mode for large composites substantially attenuated by overburden. We also revisit prior dark matter exclusions from etched mica searches, identifying shortcomings that compromise the robustness of these constraints.
\end{abstract}

\newpage

\tableofcontents

\section{Introduction}

Dark matter is known to exist through its gravitational influence on galactic rotation curves, gravitational lensing in galaxy clusters, and its imprint on density perturbations in the cosmic microwave background as well as large-scale structure. Yet its (putative) non-gravitational interactions with Standard Model matter remain undetected, and its fundamental nature is unknown. Among many dark matter candidates, composite dark matter, $i.e.$ bound states of dark sector constituents assembled through nucleosynthesis-like 
processes, phase transitions, or dissipative collapse, occupies a particularly rich and 
relatively unexplored region of theory space 
\cite{Witten:1984rs, Farhi:1984qu, DeRujula:1984axn, Wise:2014ola, Wise:2014jva,
Krnjaic:2014xza, Hardy:2014mqa, Gresham:2018anj, Gresham:2017cvl, Grabowska:2018lnd, 
Coskuner:2018are, Bai:2018dqf, Bai:2019ogh, Bramante:2018qbc, Acevedo:2020avd, 
Acevedo:2021kly, Cappiello:2020lbk,Bleau:2025klr, Picker:2025ofy,DeRocco:2025ovr }.
Just as visible matter forms atoms, nuclei, and 
macroscopic objects through the interplay of attractive forces and binding energies, 
dark sector dynamics can also produce composite states spanning 
masses and radii; for a recent review, see \cite{Bramante:2026wzh}.

For dark matter masses much above a microgram ($\sim 10^{18}\,\mathrm{GeV}$), 
conventional direct detection experiments become insensitive: the galactic 
dark matter flux through a meter-scale detector falls below one particle per year \cite{Bramante:2018qbc}. Ancient minerals provide a compelling 
alternative. A geological sample that has resided in a thermally stable, 
low-annealing environment for a billion years accumulates the equivalent of a 
gigaton-year exposure, potentially recording rare heavy dark matter transits as permanent 
tracks of lattice damage. The foundations of this approach were established in 
searches for strongly interacting massive particles and magnetic monopoles using muscovite mica \cite{Price:1986ky, SnowdenIfft:1995ke}. In those studies, cleaved mica slabs were chemically etched with hydrofluoric 
acid and inspected by optical and atomic force microscopy; the null observation of 
through-going tracks set leading exclusions on the flux of magnetic monopoles. Estimates of the dark matter scattering cross-sections excluded by these monopole track searches were subsequently obtained in \cite{Acevedo:2021tbl,Jacobs:2014yca}. A broader program of mineral-based astroparticle detection has since been proposed and explored, with theoretical frameworks developed for a range of 
geological target materials and dark matter candidates 
\cite{Baum:2018tfw,Ebadi:2021cte,Alfonso:2022meh,Baum:2023cct,Fung:2025cub,Araujo:2025rhr,Calabrese-Day:2026soq,Hedges:2026pgf}. 
A key lesson emerging from this resurgent program is that the sensitivity of any 
mineral paleodetector is ultimately governed not by exposure alone, but 
by the fidelity with which track formation thresholds and readout 
efficiencies can be established through calibration \cite{Fung:2025cub}. 
Muscovite mica is particularly well-positioned in this respect: its 
layered crystal structure permits mechanical cleaving of large slabs along 
the basal plane, its track retention over gigayear timescales is able to be 
empirically established through fission track counts, and its 
elemental composition makes it a favourable target for both heavy 
composite dark matter interactions and X-ray fluorescence readout. A further attractive feature of mica as a paleodetector is that the heavy composite signature is essentially background-free. No known Standard Model process produces a continuous, straight-line damage feature that would extend centimeters through mica sheets: spontaneous fission tracks terminate within $\sim 10~\mu$m and alpha-recoil tracks anneal on $\sim$\,Myr timescales (see Section~\ref{sec:revisiting}). A heavy composite transit is, to our knowledge, the only physical process that can produce a single, contiguous $\gtrsim$\,cm-long damage track at the depths and exposure ages relevant to this work.

In this work we take up the challenge of extending the sensitivity of muscovite mica 
as a paleodetector for heavy composite dark matter into a regime of large composites with radii from nanometers to microns where 
existing etching-based readout methods would be inefficient and where no systematic calibration of the damage mechanism has previously been performed. First, in Section \ref{sec:models} we develop a quantitative theoretical framework for 
energy deposition and melt track formation by heavy dark matter composites 
transiting mica, treating both the opaque (geometric) limit and the 
optically thin (loosely bound) limit, and deriving the expected melt track radius 
as a function of composite radius in both cases using a thermal spike framework. We validate these melt radius predictions at sub-micron scales using SRIM/TRIM 
simulations of nuclear recoil cascades, which also provide a calibration of the 
phonon efficiency factor governing the fraction of deposited energy available 
for local heating. In Section \ref{sec:readout} we demonstrate a novel readout methodology: 
rapid large-area X-ray Fluorescence (XRF) mapping using a copper backing 
contrast technique, capable of identifying micron-scale melt voids in 
cleaved mica sheets, and present a preliminary calibration using laser-ablated 
holes as track surrogates. In Section \ref{sec:sensitivity} we discuss the methodology for establishing 
and validating track retention ages in mica samples, combining 
isotope geochronology for primary crystallization age with 
in-situ background fission track counting as an internal calibration 
of the thermal annealing history. We revisit some prior assumptions about track retention age ($aka$ annealing) in mica-based heavy dark matter studies. We present prospective sensitivities 
for both opaque and diffuse composite dark matter, for benchmark mica exposures. In Section \ref{sec:conclusion}, we conclude.

\section{Heavy Composite Dark Matter and Mineral Signatures}
\label{sec:models}

We consider heavy composite dark matter states characterized by a 
total mass $M_D$ and a physical radius $R_D$, transiting through 
ancient mineral targets at a typical halo velocity $v_D \sim 
10^{-3}\,c$. These macroscopic states deposit significant integrated 
energy into geological targets along their transit path, potentially 
forming permanent damage tracks that persist over billion-year 
timescales.

The energy scale of individual nuclear scatters sets the stage for 
the resulting geological signature. For a target nucleus of mass $m_N$, 
the mean recoil energy in the elastic, heavy-composite limit ($M_D \gg 
m_N$) is
\begin{equation}
    \langle E_R \rangle \approx m_N v_D^2 \approx 20~\text{keV} 
    \left( \frac{m_N}{20~\text{GeV}} \right) 
    \left( \frac{v_D}{10^{-3}} \right)^2.
    \label{eq:erecoil}
\end{equation}
This recoil energy is roughly five orders of magnitude larger than the typical atomic binding energies ($\sim$\,eV) that anchor nuclei to their 
lattice sites in muscovite mica, and several orders of magnitude 
above the $\mathcal{O}(10\text{--}30~\text{eV})$ displacement 
thresholds discussed in the stopping-range simulations presented in Section~\ref{sec:srim}. For the 
composite dark matter considered here, the relevant question is therefore not whether primary knock-on atoms are displaced, but rather 
how much energy is deposited per unit path length, and 
whether that energy density is sufficient to produce a contiguous 
melt or vaporization track detectable by XRF mapping.

\begin{table}[]
\centering
\begin{tabular}{ll}
\hline
\textbf{Parameter}     & \textbf{Value} \\ \hline
Chemical formula       &  KAl$_2$(AlSi$_3$O$_{10}$)(OH)$_2$ \\
Density $(\rho)$            &  2.77 - 2.88 ($\sim$ 2.825)  g/cm$^3$\cite{1978usgs.rept....3R}  \\
Molar volume           & 140.716 cm$^3$/mol\cite{1978usgs.rept....3R} \\
Subterranean depth     &  20 km  \\
Temperature at 20\,km depth & 25\degree C/km of depth, $\approx 773$ K\cite{geothermalenergy}\\
Melting temperature & 1500 K \cite{1978usgs.rept....3R} \\
Specific heat capacity ($C_p$) & 816 J/kg K, at $T$ = 293 K\\
  % & 1221 J/kg/K, at T = 773 K\\
  & 1328 J/kg K, at $T$ = 1500 K \cite{1978usgs.rept....3R} \\
Thermal Conductivity ($K_d$)  &  0.6 W/K m $\left (1500~K/T\right)$\cite{Hofmeister2015}  \\
                       % &                \\
                       % &                \\
                       % &               
\end{tabular}%
\caption{Summary of muscovite mica properties used throughout this paper.
The conductivity quoted is the in-plane (parallel to the basal cleavage) value at $T=1500$~K,
scaled from the room-temperature in-plane value of $\sim 3$~W/m\,K reported
in Ref.~\cite{Hofmeister2015}; the through-cleavage value is roughly an order
of magnitude smaller and is irrelevant for the cylindrical melt-track geometry
considered here, in which heat conducts predominantly radially within the basal
plane.}
\label{tab:mica}
\end{table}

Table \ref{tab:mica} summarizes the elemental characteristics of muscovite mica used in all calculations throughout this paper, unless otherwise specified in the text.

\subsection{Composite Dark Matter Interactions: Two Models of Energy Deposition}

We investigate two limiting cases of composite dark matter 
interactions with mineral nuclei, distinguished by the optical depth 
of the composite to nuclear scattering. These two cases produce 
qualitatively different energy deposition profiles, which motivate 
the separate melt radius scalings derived in Section~\ref{sec:melt}.

\subsubsection*{Model A: The Opaque (Geometric) Limit}

This limit describes composites whose effective interaction potential 
reflects every incident nucleus within the geometric cross-section 
$\sigma_\text{geo} = \pi R_D^2$ at the halo velocity $v_D$ 
\cite{Acevedo:2021kly, Acevedo:2020avd, Wise:2014jva, 
Cappiello:2020lbk}. The energy deposition 
rate per unit path length is determined entirely by the geometric 
cylinder of rock swept out,
\begin{equation}\label{eq:Edep_opaque}
    \left(\frac{dE}{dx}\right)_\text{geo} \approx 
    n_\text{mica}\, \pi R_D^2\, \langle E_R \rangle,
\end{equation}
where $n_\text{mica} \approx 9\times10^{22}~\text{cm}^{-3}$ is the mean
nuclear number density of muscovite, computed from the density $(\rho)$ and molar
formula in Table~\ref{tab:mica}. The corresponding mean nuclei mass is $\bar m_N \approx 18~\text{GeV}$, and we adopt these as
the reference values throughout.
Since $\langle E_R\rangle \approx \bar m_N v_D^2 \simeq 20~\text{keV}$
(Eq.~\ref{eq:erecoil}, evaluated for muscovite's average nuclei mass) while
the per-nucleus enthalpy required for melting is $\mathcal{O}(0.1~\text{eV})$, opaque composites deposit energy far in excess of the melt threshold throughout the swept 
volume, driving a hypersonic cylindrical shock into the surrounding 
lattice. The resulting melt track radius is macroscopic and 
independent of the composite's internal structure, depending only on  $R_D$, as derived in Section~\ref{sec:melt}.

\subsubsection*{Model B: The Diffuse (Constituent) Limit}
This limit describes loosely-bound composites whose constituents 
interact with muscovite nuclei via a per-nucleon cross-section $\sigma_{\chi n}$, 
with the composite transparent enough that most nuclei traverse it 
without scattering \cite{Acevedo:2024lyr,Boukhtouchen:2025vvg}. We consider a loosely-bound composite of radius $R_D$ consisting 
of $N_{\chi} = M_D / m_{\chi}$ constituents, each of mass $m_{\chi}$, distributed 
uniformly with number density $n_{\chi} = 3 M_D / (4\pi R_D^3 m_{\chi})$. For coherent scattering off nuclei of mass number $A$ in 
the heavy-constituent limit, the effective constituent-nucleus 
cross-section scales as $\sigma_{\chi A} \approx \sigma_{\chi n} A^4$.

We define the optical depth of a single muscovite nucleus with mass number $A$, as the mean number of scatters received by the nucleus transiting the composite with random and uniform incidence,
\begin{equation} \label{eq:opticalDepth}
    \tau_A = 
   \frac{4}{3}\, \sigma_{\chi A}\, n_{\chi}\,  R_D \approx 
    \frac{\, \sigma_{\chi n} A^4 M_D}{\pi\, m_{\chi} R_D^2}.
\end{equation}

The mean energy deposited to the nucleus as it traverses the interior of the composite is therefore,
\begin{equation}
    \label{eq:Edep_diffuse_micro}
    E_\text{dep} = \mathcal{N}_\text{scat}\, \langle E_R \rangle = 
    \tau_A  
    m_N v_D^2 = \left( \frac{\, \sigma_{\chi n} A^4 M_D}{\pi\, m_{\chi} R_D^2}\right)m_N v_D^2.
\end{equation}
It will be more helpful to express the energy deposition in terms of the \textit{composite}-nuclei scattering cross section, which also scales as $\sigma_{D A} \approx \sigma_{n} A^4$.
As the dark matter traverses a length $L$ through a cloud of nuclei with mass number $A$ with number density $n_A$, it will scatter against nuclei $N = \pi R_D^2\,\tau_A\,  n_A L$ times. But one can also write the number of scatters as $N = \sigma_{D A}\, n_A\, L$. Comparing these two equations, we write the optical depth as the ratio of the total composite-nuclei interaction cross section with the geometric cross section,
\begin{equation}\label{eq:opticalDepth2}
    \tau_A = \frac{\sigma_{DA}}{\pi R_D^2}.
\end{equation}
Comparing with Equation \ref{eq:opticalDepth}, we recover the expected $\sigma_{DA} = N_\chi \sigma_{\chi A}$ and $\sigma_{n} = N_\chi  \sigma_{\chi n}$.
The mean energy deposition per nuclei is then,
\begin{equation}
    \label{eq:Edep_diffuse_macro}
    E_\text{dep}  = \left(\frac{\sigma_{DA}}{\pi R_D^2}\right)m_N v_D^2.
\end{equation}
Unlike the opaque (geometric) case, $E_\text{dep}$ in this regime depends on the
micro-physical parameters of the composite — equivalently $\sigma_{DA}$
together with $R_D$, or $(\sigma_n, m_\chi, M_D, R_D)$ in the per-nucleon
parametrization. This allows the deposited energy per site to range from well below to well above the melt threshold as the composite's column
density is varied. In this work we focus on the parameter space where $E_\text{dep}$ is sufficient to produce a detectable melt track, which essentially
maps onto the intermediate and high-$\beta$ regimes of
Section~\ref{sec:melt}.

\subsection{Melt Radius: The Thermal Spike Model}
\label{sec:melt}
Both models above deliver energy to the rock on a timescale set by the
composite transit time, $t_\text{dep} \sim R_D / v_D$. Here $R_D$ is
the transverse scale of the prompt energy deposition: in the opaque limit
this is the geometric radius of the composite, and in the diffuse limit it is the chord traversed by a given nucleus through the composite interior, with $R_D$ setting the outer scale. For our largest composites of interest, $R_D \sim 100~\mu\text{m}$,
\begin{equation}
    t_\text{dep} \approx 3.3 \times 10^{-10}~\text{s} 
    \left(\frac{R_D}{100~\mu\text{m}}\right).
\end{equation}
This should be compared to the timescale for the deposited heat to 
diffuse radially outward from the track. Using the high-temperature ($T = 1500$~K) thermal properties of muscovite
mica listed in Table~\ref{tab:mica}~\cite{Hofmeister2015}, the thermal
diffusivity is $D = K_d / \rho C_p \approx 1.6\times 10^{-7}~\text{m}^2/\text{s}$. The diffusion 
timescale over a length scale $L$ is then, 
\begin{equation}
    t_\text{diff} \approx 
    \frac{L^2}{D} \approx 
    6.25\times10^{-2}~\text{s}~
    \left(\frac{L}{100~\mu\text{m}}\right)^2
    \left(\frac{T}{1500~\text{K}}\right).
\end{equation}

Since $t_\text{dep} / t_\text{diff} \sim 10^{-8}$ across 
all length scales relevant to this work, the energy deposition is  effectively instantaneous on the thermal diffusion timescale. We are 
therefore justified in treating the track as an adiabatic energy injection. In the following we apply the Sedov-Taylor thermal spike approximation \cite{Sedov1959, Zeldovich1967} to derive the expected melt radius as a function of $R_D$ for both composite models.

In the opaque geometric limit, a macroscopic dark matter state interacts by displacing all target nuclei within its cross-sectional area $\pi R_D^2$. The total kinetic energy deposited per unit path length is $dE/dx \approx \rho v_D^2 \pi R_D^2$, where $\rho = n_{\rm mica} \bar{m}_N$ is the mean target mass density. However, electronic stopping partitions a fraction of this energy into ionization, which can diffuse over large radial distances before thermalizing. We parameterize the localized heat deposition via a phonon efficiency factor, $\eta$, representing the fraction of energy transferred strictly to nuclear cascades and phonon energy.

To form a contiguous melt track of radius $R_{\mathrm{melt}}$, this prompt thermal energy must raise the enclosed cylindrical volume to the mineral's melting point and provide the latent heat of fusion. This requires an energy per unit length $dE_{\rm req}/dx \approx \rho (C_p \Delta T_{melt} + H_f) \pi R_{\mathrm{melt}}^2$, where $C_p$ is the mineral specific heat capacity, $\Delta T_{melt}$ is the temperature increment to the melting point and $H_f$ is the specific latent heat for melting the mica. Equating the effective heat deposited, $\eta (dE/dx)$, to $dE_{\rm req}/dx$ yields an adiabatic scaling melt radius \cite{Sedov1959} that is independent of the target density
\begin{equation}
    R_{\mathrm{melt}} \approx R_D \sqrt{\frac{\eta v_D^2}{C_p \Delta T_{\mathrm{melt}} + H_f}}.
    \label{eq:Rmelt_adiabatic}
\end{equation}

For muscovite mica, using the melting temperature $T_{\rm melt} = 1500$\,K, specific heat capacity $C_p$ at $T = 293$~K from Table~\ref{tab:mica}, and using a typical specific latent heat of fusion $H_f \approx 4 \times 10^5$~J/kg, this evaluates to a macroscopic track radius of $R_{\mathrm{melt}} \approx 255 \sqrt{\eta} R_D$. For standard low-velocity nuclear recoils in insulators where $\eta \approx 0.7$ \cite{LindhardScharffSchiott1963,Toulemonde2000}, the melt track scales with dark matter composite radius as $R_{\mathrm{melt}} \approx 220 R_D$. Note that simple geothermal models predict a temperature of $\sim 773$~K at $20$~km depth \cite{geothermalenergy}; however, validating that a given mineral sample has remained at such a depth throughout its lifetime is non-trivial, and some cratonic regions in fact sustain temperatures as low as $\sim 600$~K even at $20$~km \cite{majorowicz2019thermal}. We therefore conservatively adopt an initial temperature of $T_i = 293$~K, and hence, $\Delta T_{\rm melt} = (1500 - 293)$~K $\approx$ 1200~K when computing the energy required to produce a melt track.

One might assume that thermal diffusion introduces a substantial efficiency penalty to this idealized limit. In standard single-ion track formation, heat conducts outward with a Gaussian profile, stranding a substantial portion of the deposited energy near the origin. However, a macroscopic projectile travelling at $v_D \approx 300$~km/s induces a local Mach number $\mathcal{M} \gg 1$. Under these conditions, the target lattice yields entirely to kinetic pressure, and the energy deposition drives a hypersonic cylindrical shock wave. In this Sedov-Taylor blast regime, energy is efficiently swept radially outward by the shock front \cite{Sedov1959, Zeldovich1967}. 

This can be compared with an optically thin composite, where scattering is distributed throughout the composite interior with an optical depth $\tau_A \ll 1$. Because the dark matter transit time is many orders of magnitude shorter than the lattice thermal diffusion timescale, the energy deposition is effectively instantaneous. The local deposited energy to a single nucleus of mass number $A$ follows the projected chord length of the spherical composite, yielding a centrally peaked initial energy profile,
\begin{equation} \label{eq:energyProfile}
    \Delta \epsilon(r) = \frac{3}{2} \eta \,m_A\tau_A v_D^2 \sqrt{1 - \left(\frac{r}{R_D}\right)^2},
\end{equation}
where we have assumed the mass of the constituent of the dark matter composite is larger than the nuclear mass. The total energy per unit length deposited into the mica inside an inner radius $r<R_D$ of the composite is then,
\begin{equation}\label{eq:innerRadiusEnergyDep}
    \Delta \epsilon(<r) = \pi \eta \tau_{\mathrm{mica}}\rho R_D^2 v_D^2 \left(1 - \left(1-\left(\frac{r}{R_D}\right)^2\right)^{3/2}\right),
\end{equation}
where $\tau_{\mathrm{mica}}$ is the optical depth averaged over the mass of nuclides in the mica so that, if $f_{A}$ is the mass abundance of the nuclide $A$ in the mica, then $\tau_{\mathrm{mica}}=\sum_{A \in \mathrm{mica}}\tau_A f_{A}$.
To melt a track of radius $r<R_D$, this must be greater than the melting energy per unit length $\rho (C_p \Delta T_{melt} + H_f) \pi r^2$. We define the dimensionless energy ratio $\beta \equiv \pi\eta \tau_{\mathrm{mica}} v_D^2 / (C_p \Delta T_{\mathrm{melt}} + H_f)$, which determines to what extent the composite melts the mineral in its interior. 

If $\beta\leq\frac{2}{3}$, the deposited energy never exceeds the local energy required, even for nuclei passing through the centre of the composite, and there is no melt track. If $\frac{2}{3}<\beta \leq 1$, the dark matter melts a track of radius,

\begin{equation}\label{eq:lowMeltRadius}
    R_{\mathrm{melt}} \approx R_D \sqrt{\frac{3\beta^2-1-(1-\beta)^{\frac{3}{2}}\sqrt{3\beta+1}}{2\beta^2}}\; \text{for } \frac{2}{3}<\beta \leq 1,
\end{equation}
where the melt size monotonically increases from $R_{\mathrm{melt}}=0$ at $\beta=\frac{2}{3}$ to $R_{\mathrm{melt}}=R_D$ at $\beta=1$. Conversely, in the high-melt regime ($\beta > 1$), the energy deposited inside $R_D$ is enough to melt a track with radius $R_{\mathrm{melt}}> R_D$. The superheated interior drives a radial shockwave outward into the unperturbed lattice. In this macroscopic limit, the specific internal geometry of the composite becomes irrelevant, and the system reduces effectively to the opaque geometric composite detailed above, where a super-melt temperature shock propagates outward:
\begin{equation}\label{eq:highMeltRadius}
    R_{\mathrm{melt}} \approx R_D \sqrt{\beta} =  R_D \sqrt{\frac{\eta \tau_{\mathrm{mica}} v_D^2}{C_p \Delta T_{\mathrm{melt}} + H_f}} \quad \text{for } \beta > 1.
\end{equation}

\subsection{SRIM Simulation and Calibration}
\label{sec:srim}
The analytic scaling above relies on a clean separation between energy
deposition and thermal diffusion, an approximation we have justified on
purely macroscopic grounds. To validate it at sub-micron composite radii,
where individual recoil cascades are still resolvable rather than
hydrodynamic, we turn to numerical SRIM/TRIM simulations of the cascades
themselves. These simulations also calibrate the phonon efficiency factor
$\eta$ that enters Eqs.~\ref{eq:Rmelt_adiabatic}, \ref{eq:lowMeltRadius},
and~\ref{eq:highMeltRadius}.

While traversing a mineral, a composite will interact with some number of nuclei in the mineral's crystal lattice. These "primary knock-on atoms" (PKAs) will, if imparted sufficient energy, break out of their bonds in the lattice, recoil away and themselves cause a secondary "cascade" of displaced and colliding SM nuclei. We model these cascades using the SRIM (Stopping and Range of Ions in Matter) \cite{ZIEGLER20101818} simulation, which yields the energy deposition within the target material from nuclear recoils, electronic stopping, and phonon production. We use these SRIM simulations to characterize melt track radii for composites with sub-10 nm radii; in addition, they are a useful calibration tool for the phonon efficiency factor $\eta$. In this analysis, we also use the \texttt{pysrim} package for Python, which acts as a wrapper for SRIM and allows for automation of SRIM calculations and simplifies the parsing of the output data \cite{Ostrouchov2018}.

A SRIM simulation is initialized by defining the stoichiometric composition of the target material, its density, as well as the displacement energy and lattice binding energy of each of the elements in the target. The displacement energy is defined as the energy a recoiling nucleus needs in order to leave its lattice site and move an atomic spacing away from it; i.e. the energy required to overcome the lattice forces. If a nuclear scattering event does not impart sufficient energy to meet this displacement energy threshold, no displacement will occur and the recoiling atom will release its kinetic energy as phonons as it returns to its original site. The lattice binding energy simply defines the energy lost by an atom leaving its lattice site to binding forces in the lattice. 

\begin{figure}
    \centering
    \includegraphics[width=\textwidth]{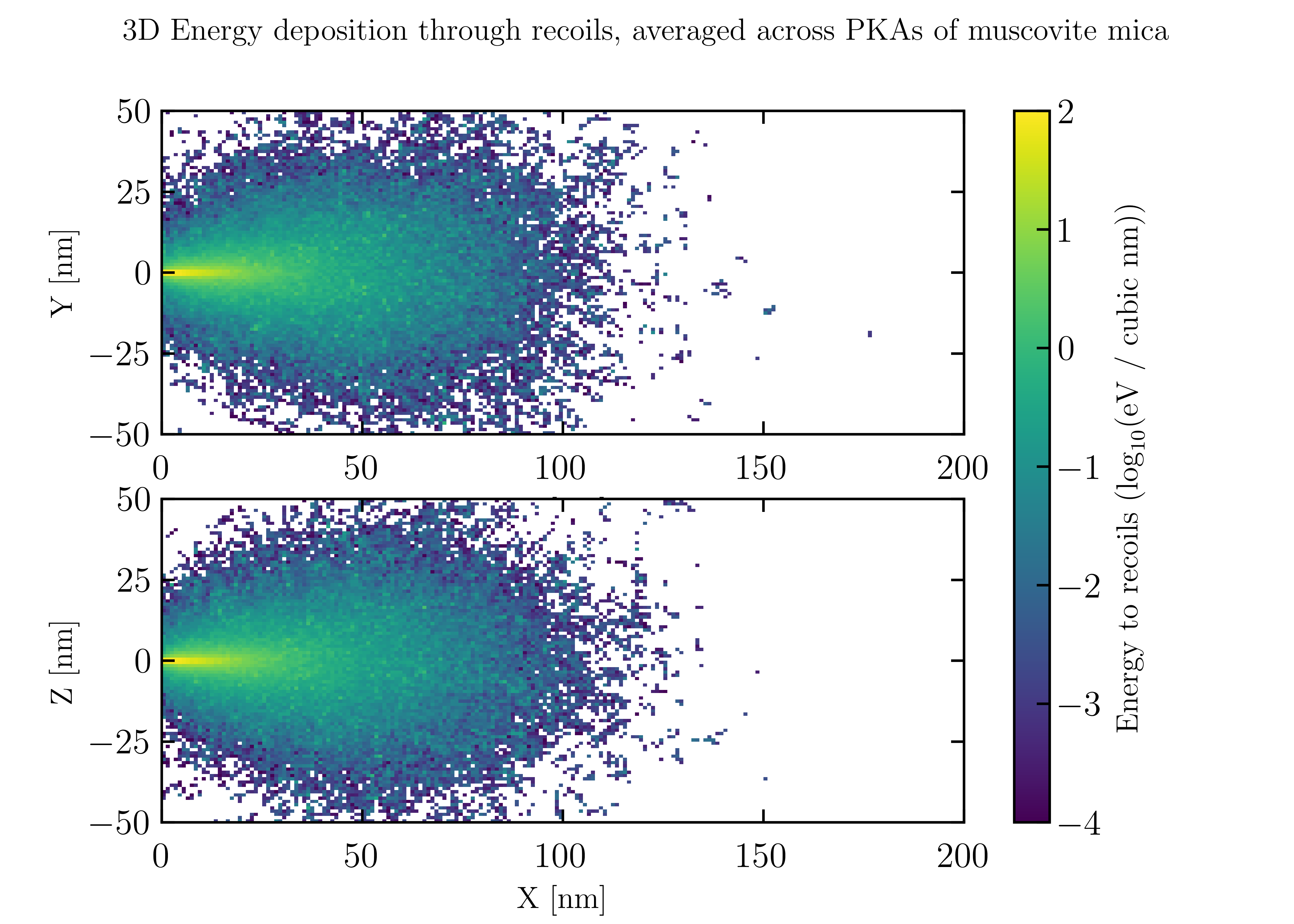}
    \caption{Average energy deposited due to a recoil cascade induced by a single PKA from a composite transit through muscovite mica. The PKA initially travels in the x-direction (the depth). Top: energy deposition in X-Y direction, slicing at $z = 0$. Bottom: energy deposition in X-Z direction, slicing at $y = 0$.}
    \label{fig:SinglePKA}
\end{figure}

\begin{figure}
    \centering
    \includegraphics[width=\linewidth]{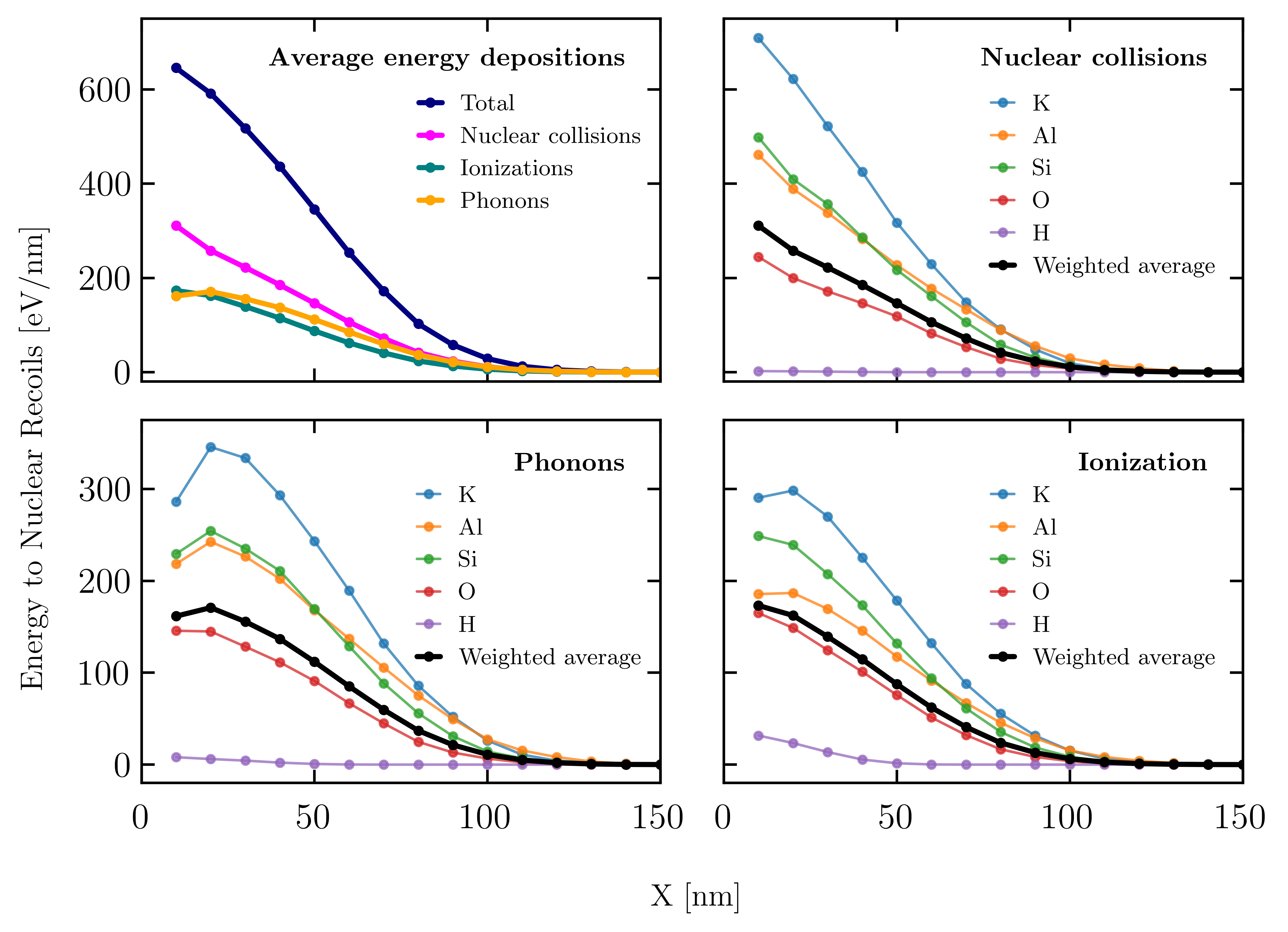}
    \caption{Energy depositions due to a recoil cascade induced by a single primary knock-on atom from a composite transit through muscovite mica. The depth, X, is defined as the initial direction of travel of the PKA. \textit{Top left:} A comparison of the average energy depositions per unit length to nuclear collisions, ionizations, and phonons. \textit{Top right:} Energy deposition through nuclear collisions. \textit{Bottom left:} Energy deposition through phonon production. \textit{Bottom right:} Energy deposition through ionization.}
    \label{fig:Edep1D}
\end{figure}

The incident Standard Model ion and its initial kinetic energy are the other two inputs. For each element in muscovite mica, we determine the distribution of energies imparted to it from scattering with an incoming dark matter composite with speed $v_D$, from the hard-sphere approximation,
\begin{equation}
    E_{dep} = \frac{2 m_A M_D (1 - \cos{\theta_{cm}})}{(m_A + M_D)^2}\times \frac{1}{2} M_D v_D^2.
    \label{eq:Edep}
\end{equation}
Using a distribution of energies as input energies for each element in SRIM, along with the stoichiometric abundance of the elements, we can determine the average energy deposition expected in muscovite mica due to a single PKA. The initial dominant energy deposition occurs through nuclear scattering throughout the recoil cascade. A three-dimensional map of the nuclear energy deposition for a single PKA is shown in Figure \ref{fig:SinglePKA}. where the x-direction (or depth) is parallel to the initial velocity of the PKA. We also plot the one-dimensional nuclear energy deposition as a function of depth in the top right panel of Figure \ref{fig:Edep1D}. 

As recoiling ions lose energy, they are no longer able to dislodge other atoms through scattering. Instead, the kinetic energy imparted to the scattered atoms is transferred to the crystal lattice through phonons, as shown in the bottom left panel of Figure \ref{fig:Edep1D}, and this energy contributes to heating the crystal, disrupting its structure and, at a sufficiently high temperature, melting the crystal. On the other hand, a recoiling ion also loses energy to electrons in the crystal, as they are excited to higher energy states or ionized completely. The energy deposition through these ``electronic stopping'' processes are shown in the bottom right panel of Figure \ref{fig:Edep1D}. The energy transfer between these energetic electrons and the crystal lattice is governed by the electron-phonon coupling, which varies across materials but is generally inefficient~\cite{Toulemonde2000,KaganovLifshitzTanatarov1957}.
 Conservatively, we assume that this process is completely inefficient, such that none of the energy deposited to electrons is transferred to the crystal lattice, and only the energy deposited through nuclear scattering and phonons contributes to heating the crystal. We can then estimate the efficiency factor $\eta$ for muscovite mica using our SRIM simulations to be simply,

\begin{equation}
    \eta = \frac{E_{\rm nuc} + E_{\rm phonon}}{E_{\rm nuc} + E_{\rm phonon} + E_{\rm elec}},
    \label{eq:eta}
\end{equation}
which we find to be about $\eta \simeq 0.75$.

We also determine the expected melt radius from SRIM and compare this to our analytic results from Section \ref{sec:models}, focusing in particular on the geometric scattering regime. Using the one-dimensional nuclear scattering and phonon energy deposition per PKA, in Fig. \ref{fig:Edep1D}, we determine the total energy deposition due to a single composite of radius $R_D$. We assume that the PKAs are uniformly distributed across the composite cross-section and emerge radially outward from its surface. Under these assumptions, the depth X maps directly onto the distance from the composite centre $r$, and the expected energy deposition per unit volume is,

\begin{equation}
    \frac{dE(r)}{dV} = \frac{N_{\rm PKA}~f_{\rm PKA}(r)}{2\pi r} = \frac{n_{mica}~\pi R_D^2~f_{\rm PKA}(r)}{2\pi r},
    \label{eq:1dEnergy}
\end{equation}
where $f_{\rm PKA}(r) = f_{\rm PKA,~nuc}(r) + f_{\rm PKA,~phonon}(r)$ is the energy deposition per PKA per unit length described in Fig. \ref{fig:Edep1D}. 

The energy deposition per unit volume is more accurately modelled using the three-dimensional collision data given by SRIM, and shown previously in Fig. \ref{fig:SinglePKA}, rather than a one-dimensional estimate. If we once again assume that PKAs are distributed evenly and recoiling radially outwards from the composite surface, we need only to translate and rotate the three-dimensional energy deposition to match each PKA's initial location, and combine to form a total energy deposition. However, this can be a computationally-intensive endeavour as the number of PKAs increases with $R_D^2$, so it is useful to characterize the agreement between these two methods by comparing their predicted melt radii. 

One caveat is that SRIM only outputs phonon and ionization energy deposition data as a one-dimensional function of depth, and only nuclear collisions are saved in three dimensions. We approximate the three-dimensional distribution for the phonon energy deposition by assuming that the energy is distributed similarly to the nuclear scatters,
\begin{equation}
    \frac{dE(x, y,z)}{dV} = \frac{N_{\rm scatters} (x,y,z)}{\int_{y,z}N_{\rm scatters} (x,y,z)} f_{\rm PKA,~phonon}(x).
    \label{eq:3dEnergy}
\end{equation}
This assumption is not completely accurate, as phonons are produced from low-energy scatters that do not induce a displacement of the target, and are therefore more likely to be produced towards the tail end of a nuclear scatter cascade. As we are only aiming to compare these three-dimensional results with the one-dimensional case, however, this simplification is acceptable.

\begin{figure}
    \centering
    \includegraphics[width=\linewidth]{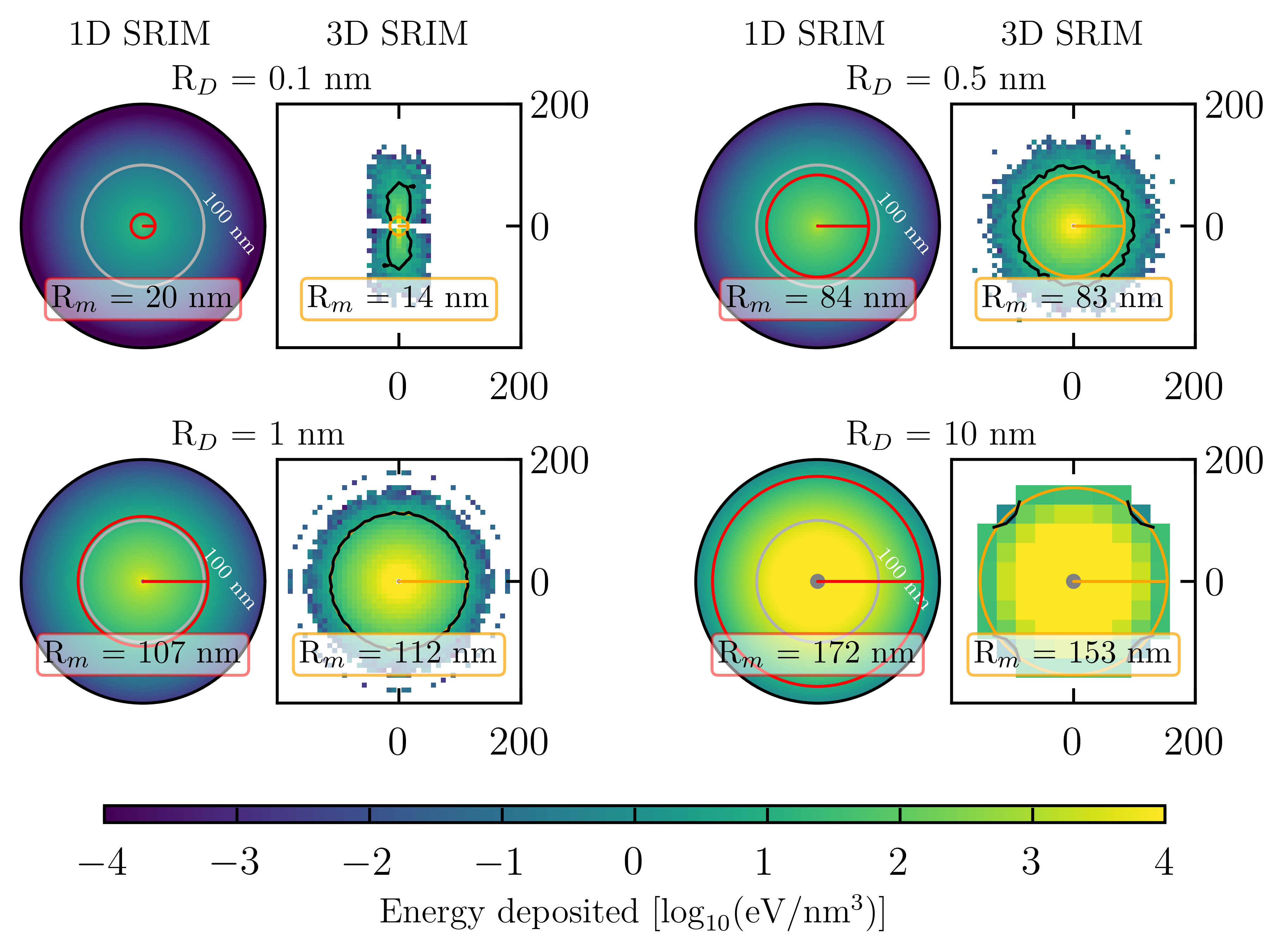}
    \caption{Total energy deposition per cubic nanometre, for composites of varying $R_D$. For each $R_D$, the "1D SRIM" column shows the energy deposition calculated from Eq. \ref{eq:1dEnergy}, and the "3D SRIM" column is calculated from Eq. \ref{eq:3dEnergy}. The grey shaded circles in the centre of each plot represent a cross-sectional view of the dark matter composite, which is travelling into the page. The region enclosed in the red and orange circles is expected to be sufficiently heated to be melted, and the corresponding melt radius $R_{\rm melt}$ is labelled. For $R_D \gtrsim 1$~nm, SRIM's assumption of an unchanged crystal lattice breaks down, and the cascade picture begins to underestimate the melt radius relative to the geometric estimate; see the text of Section~\ref{sec:srim} for details. The bottom row of this figure shows the onset of this discrepancy.} 
    \label{fig:srim}
\end{figure}

In Fig. \ref{fig:srim}, we show a cross-sectional view of total energy deposition per cubic nanometre, for composites of varying $R_D$. For each, we show the one-dimensional energy deposition from Eq. \ref{eq:1dEnergy}, juxtaposed with the three-dimensional energy deposition from Eq. \ref{eq:3dEnergy}. The grey shaded circles in the centre of each plot represent a cross-sectional view of the dark matter composite, which is travelling into the page. The region enclosed in the red and orange circles is expected to be sufficiently heated to be melted, and the corresponding melt radius $R_{\rm melt}$ is labelled. 

\begin{figure}
    \centering
    \includegraphics[ width=\textwidth]{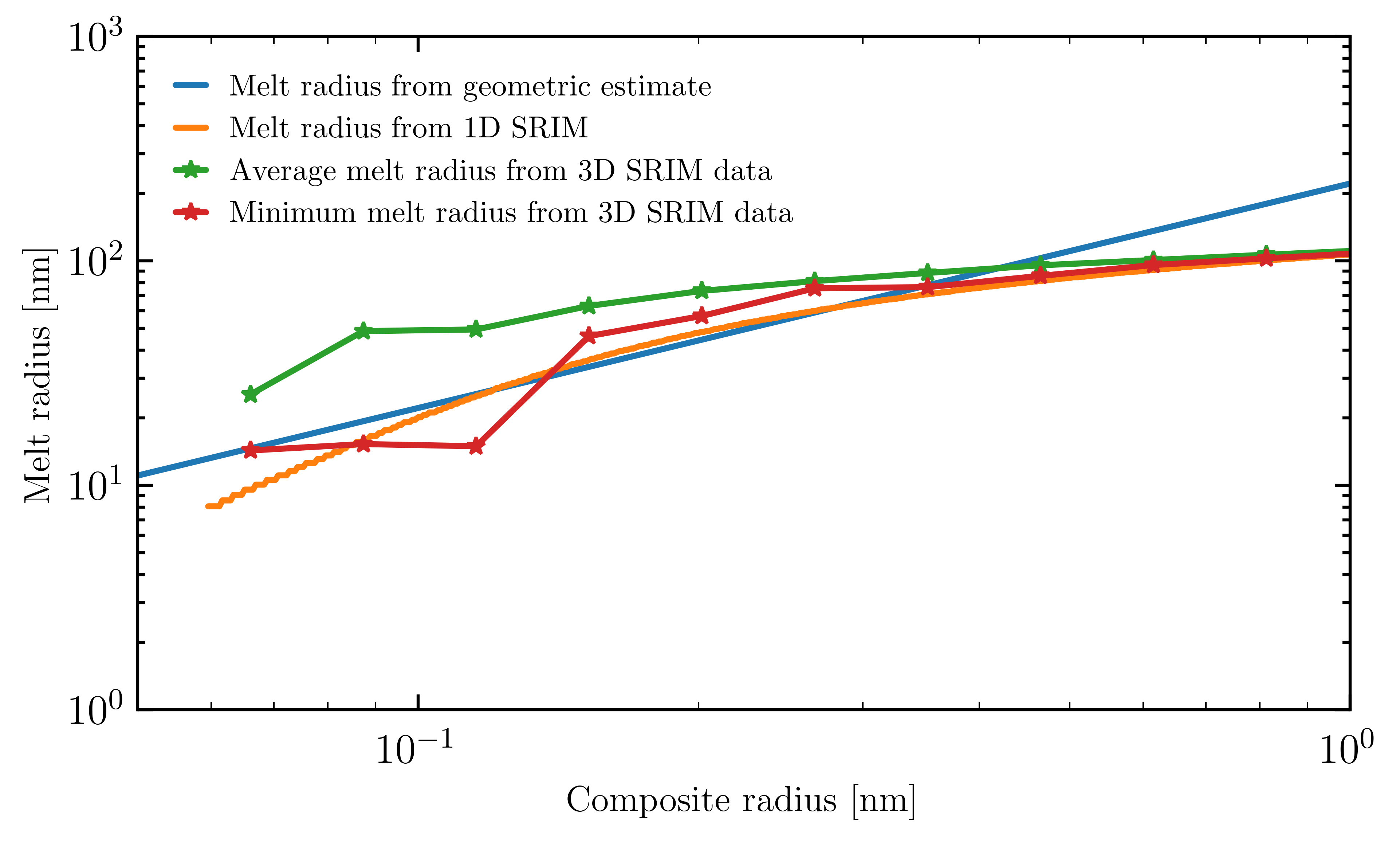}
    \caption{A comparison of the predicted melt radius for varying composite radii in the geometric scattering regime. The orange line shows the melt radius expected from the energy deposition determined by SRIM, while the blue line shows the melt radius from Eq. \ref{eq:Rmelt_adiabatic}. The remaining curves trace the melt radius extracted from the 3D SRIM energy-deposition distribution shown in Figure~\ref{fig:srim}, for the sub-micron range over which the SRIM cascade picture remains tractable.} 
    \label{fig:srimcalibration}
\end{figure}

In Fig. \ref{fig:srimcalibration} we compare these melt radius results with those calculated analytically in Eq. \ref{eq:Rmelt_adiabatic}. It is evident that the results are in good agreement over the range $R_D \sim 0.1 - 1$ nm, corresponding to melt radii of $20 - 200$ nm. We also truncate the melt radius calculations at $R_D < 0.05$ nm, where we expect a composite to cross less than 1 target nucleus per nanometre travelled into the material, based on the number density of muscovite mica. We expect these composites to induce discrete, individual scattering defects at intervals along their trajectory, rather than one cylindrical melt track, thus we do not focus on this phenomenological space in this paper.

As the composite radius increases, the number of recoiling atoms increases as $R_D^2$, and the number of defects induced in the crystal lattice also increases. At $R_D = 1$ nm, and using the vacancy production rate from SRIM simulations, we expect up to $50\%$ of the atoms in the crystal to be displaced from their lattice sites in the central region of the cascade near the composite boundary. The underlying assumption of an unchanged crystal lattice which the SRIM simulation software makes no longer holds, and as SRIM cannot take into account the bulk motion of recoiling atoms outwards as they cascade, it underestimates the melt radius as $R_D$ increases. The behaviour shown in Figure \ref{fig:srimcalibration} is thus as expected: the SRIM calculations validate the geometric estimate within the region where their assumptions hold. 

The proposed XRF mineral readout method, which will be discussed in more detail in Section \ref{sec:readout}, is currently sensitive to melt tracks with radii greater than 25 $\mu$m, corresponding to composites with $R_D \sim 0.1~\mu$m. These composites are too large for their energy deposition to be modelled through SRIM, and hence, the geometric estimate will be used to calculate damage size throughout the rest of this paper, along with the calibrated efficiency factor $\eta = 0.75$ from the SRIM analysis.

\section{Experimental Readout of Mica with XRF Transmission}
\label{sec:readout}
To identify micron-scale damage tracks from dark matter induced nuclear recoils, we employ X-ray fluorescence (XRF) mapping. This non-invasive technique presents a novel and unique opportunity to determine a material’s elemental composition. In XRF, the sample is irradiated with high-energy photons produced by an X-ray tube (e.g., a heated cathode and Rh anode generating bremsstrahlung and characteristic radiation). 
These photons eject inner-shell electrons via the photoelectric effect, and the resulting vacancies are filled by electrons from higher energy shells, emitting a spectrum of X-ray photons unique to each element. The shell labels K,L and M correspond to principal quantum number $n$ = 1,2, and 3 respectively; this naming convention predates quantum mechanics, introduced empirically by Siegbahn before the underlying orbital structure was understood \cite{Jenkins1991}. In particular, K$\alpha$ and K$\beta$ emissions arise when a K-shell (1s) vacancy is filled by a 2p (L-shell) or 3p (M-shell) electron respectively. In the setup we will employ here, a spectrum of X-ray energy versus counts is generated at a locus of points on a scanned plane. From this we reconstruct the scanned area's X-ray response and compare with optical imaging to determine the spatial distribution of elements within the sample. Using this mapping approach, we aim to identify localized melt features associated with damage tracks, by their X-ray emission.

\subsection{Setup: The Copper Backing Method}

A thin Cu sheet is placed beneath the mica sample. Any damage track that penetrates the crystalline layers increases the exposure of the underlying Cu to the incident X-ray beam. In contrast, intact regions of mica attenuate a portion of the incident and emitted X-ray photons, reducing the observed Cu signal. As a result, melt regions in the mica sample produce localized Cu fluorescence intensities that can be seen on the XRF map. Cu was selected for several reasons; First, Cu is not naturally present in mica, a silicate mineral containing the elements K, Al, Si, Mg, and Fe. The appearance of Cu fluorescence transmitted through the mica therefore provides a tracer of the mica substrate structure. Second, Cu produces strong characteristic X-ray emission lines at approximately 8.04 keV (K$\alpha$,K-L3) and 8.90 keV (K$\beta$,K-M3), corresponding to 2p$\rightarrow$1s and 3p$\rightarrow$1s transitions respectively \cite{Jenkins1991}. Finally, higher-Z elements such as Cu generate stronger fluorescent signals than low-Z elements, which can otherwise be difficult to resolve due to background scattering of X-rays in air. 

To calibrate the copper backing method for damage features intended to mimic nuclear recoil damage from dark matter, we prepare the sample by mechanically cleaving the mica to produce a thin, even surface. Then, we laser-ablated two melt regions of diameter $\sim 150~\mu$m and $\sim 50~\mu$m to serve as dark matter damage proxies. The resulting features can be seen in the middle (optical) panel of Figure~\ref{fig:xraycalibration}, with the corresponding XRF response shown in the right panel.

To readout the mica sample using X-ray fluorescence, first plexiglass supports were placed at the edges of the sample to maintain contact with the Cu sheet and to suppress warping over the scan area. The plexiglass does not obscure any region of interest and contributes a negligible fluorescence signal due to its low-Z chemical composition (primarily C, H, and O), which produces no characteristic emission lines in the energy range relevant to this measurement.

The distance between the X-ray emission head and the sample, for the scans discussed in this paper, was reported by the M6 Jetstream as $f = 56.12$ mm. This implies an X-ray emission spot size of $s = 36$ $\mu$m on the mica sample. If the sample surface deviates from the focal plane by $\Delta z$, the effective spot size would broaden, following $s_{\text{eff}} = s + 2\Delta z\, \alpha$ ref.\cite{XRF:poly}, where $\alpha$ is the output divergence half-angle of the capillary channel that focuses the beam before leaving the X-ray emission head.  The effective spot size $s_{\text{eff}}$ is approximated using $\alpha \approx 1.3\theta_c$, where $\theta_c = (30\ \text{keV}/E)$ mrad is the critical angle for total internal reflection within the capillary channel \cite{XRF:poly}. 

In this work, the sample thickness is 0.1 mm, bounding the maximum surface deviation to $\Delta z \leq 0.1$ mm. This yields a maximum spot size variance of $\epsilon \equiv 2\Delta z\, \alpha/s \approx$ 5\%, establishing a flatness tolerance of 0.1 mm. Macroscopic sample warping therefore does not constitute a dominant limitation on spatial resolution for the sample thicknesses considered. When the sample area lies within the flatness tolerance, the spatial resolution of the XRF map is instead determined by the minimum feature size that produces a statistically significant difference in Cu K$\alpha$ intensity between melt feature pixel and an intact mica pixel. These criteria ensure that the XRF signal can be reliably mapped to the corresponding optical image. Table~\ref{tab:micaparams} summarizes the key operating parameters used in this work.

\begin{table}[t]
\centering
\begin{tabular}{ll}
\hline
\textbf{Parameter} & \textbf{Value} \\
\hline
X-ray tube & Rh anode \\
Accelerating voltage & 50 kV \\
Tube current & 350 $\mu$A \\
Beam spot size & 36 $\mu$m \\
Step size & 30 $\mu$m \\
Dwell time & 25 ms /pixel \\
Detector & Silicon drift detector (SDD) \\
Energy resolution & FWHM $<$ 145 eV at Mn K$\alpha$ \\
Filter & Al 12.5 $\mu$m\\
Scan area & 8.5 $\times$ 4.5 mm \\
Scan duration & 18 minutes \\
\hline
\end{tabular}
\caption{Operating parameters for the Bruker M6 Jetstream used in all scans reported in this work. Energy resolution is quoted at the Mn K$\alpha$ line (5.898 keV), the standard reference for SDD characterization. The nearest mica constituent emission line to Cu K$\alpha$ (8.048 keV) is Fe K$\alpha$ at 6.404 keV, a separation of 1.644 keV, well above the detector energy resolution threshold. An aluminum filter is placed in the beam path to attenuate low-energy photons and reduce background in the low-energy region of the spectrum.}
\label{tab:micaparams}
\end{table}

%\subsection{Contrast Mechanism}
The detection of a dark matter induced recoil track relies on the spatial variation of Cu K$\alpha$ intensity across the scanned surface. At each pixel, the measured signal reflects the degree to which the incident X-ray beam is attenuated by the mica before reaching the Cu sheet and travelling to the detector. The transmitted Cu K$\alpha$ intensity through an intact region of mica of thickness $t$, follows the Beer-Lambert law:

\begin{equation}
    I_{\text{mica}} = I_0\, e^{-\mu t}
    \label{eq:beerlam}
\end{equation}

\noindent where $I_0$ is the incident intensity at the sample surface and $\mu = (\mu/\rho)\rho = 124\ \mathrm{cm}^{-1}$ is the linear attenuation coefficient of mica at the Cu K$\alpha$ energy of 8.04 keV. This is estimated from the mass attenuation coefficient $\mu/\rho = 43.9\ \mathrm{cm}^{2}\ \mathrm{g}^{-1}$ and density $\rho = 2.825\ \mathrm{g\ cm}^{-3}$ \cite{NIST:atten}. For a sample thickness of $\sim$100$\mu$m, the transmittance is $T_{mica} = I_{mica}/I_0= 0.29$, meaning this fraction of Cu K$\alpha$ photons are expected to pass through intact mica of this thickness. 

\subsection{Calibration and Efficiency}

\begin{figure}[h!]
    \centering
    \includegraphics[width=0.9\linewidth]{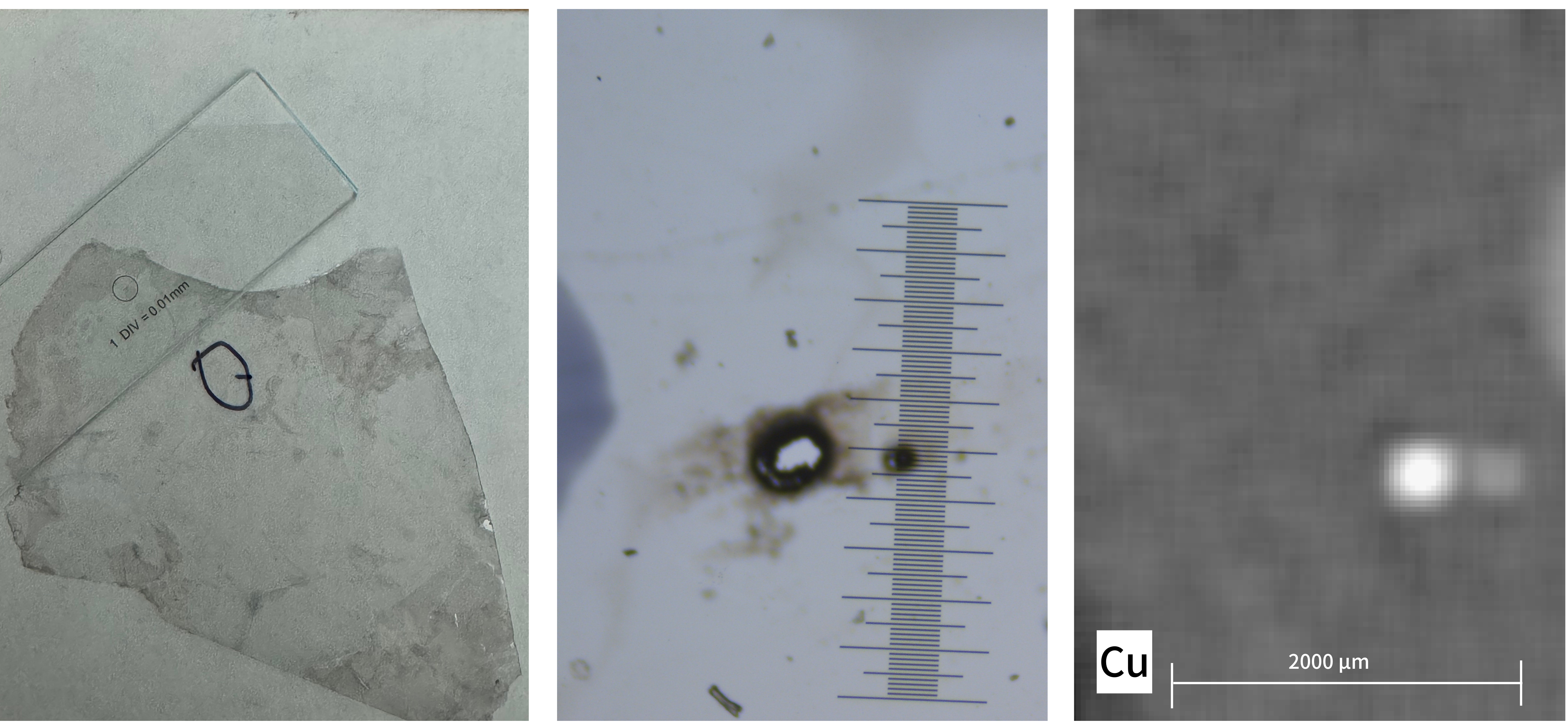}
    \caption{From left to right, three views of the same calibration sample.
\textit{Left:} wide-field photograph of the cleaved muscovite sheet, with the two laser-ablated calibration spots circled.
\textit{Middle:} optical microscopy of the marked region (Opti-Tech
microscope + BenQ camera) showing the laser-ablated melt features of
$\sim 150$ and $\sim 50~\mu$m diameter; a 100~$\mu$m ruler with
10~$\mu$m divisions is overlaid for scale.
\textit{Right:} Cu~K$\alpha$ XRF elemental map of the same
$0.2~\mathrm{cm}^{2}$ region, on which each laser feature manifests as a
localized enhancement of the characteristic Cu~K$\alpha$ emission. Detection
of the $50~\mu$m feature in the XRF map establishes a minimum detectable
track radius of $R_{\text{min}} = 25~\mu$m under the operating conditions
of this work.}
\label{fig:xraycalibration}
\end{figure}

The 150 $\mu$m and 50 $\mu$m melt features were produced in the mica sample by laser ablation. The 50 $\mu$m feature represents the lower bound of the calibration range, which establishes a minimum detectable feature radiusof 25 $\mu$m, under the operating conditions of this work, while the 150 $\mu$m feature provides a higher-contrast reference.

The mica sample was scanned over an area of $8 \times 4.5$ mm with step size of $\Delta x = 30\ \mu$m and a dwell time of $t_{\text{dwell}} = 25$ ms pixel$^{-1}$, and a total scan time of 18 minutes (Table \ref{tab:micaparams}). This results in a total scan area of 38.25 $\text{mm}^2$ divided into 42 450 pixels. At our dwell time this is 46 minutes per $\text{cm}^2$. Both melt regions were identified in the Cu K$\alpha$ elemental map as localized enhancements in count rate relative to the intact mica background. The Cu K$\alpha$ count rate for the 50 $\mu$m melt region was 85596 gross counts, compared to 74005 gross counts in intact mica, giving a fractional difference of $14\%$, i.e.\ a measured contrast of
$C_{\text{measured}} = 1 - I_{\text{intact}}/I_{\text{50 $\mu$}} \approx 0.14$.
This is well below the $C =1-T_{mica} \approx 0.71$ contrast predicted by the Beer-Lambert model if the 50 $\mu m$ region were a complete void, indicating that the larger laser-ablated melt region contains ample partially re-solidified material; the feature is nonetheless readily detectable as a localized Cu~K$\alpha$ enhancement in the XRF map. 

These calibration results, shown in Figure \ref{fig:xraycalibration}, establish that the primary practical constraint for scaling the XRF transmission method to the full target exposure area, will be the total scan time. Reduction of the total scan time might be achieved by increasing the step size $\Delta x$, or reducing the dwell time $t_{\text{dwell}}$, provided the count rate remains sufficient for statistically significant detection at each pixel.

\section{Searching for Composite Dark Matter in Melted Mica}

With the energy-deposition models of Section~\ref{sec:models} and the XRF
readout calibration of Section~\ref{sec:readout} in hand, we combine the
geological, astrophysical, and detector inputs to derive a projected dark
matter sensitivity. Section~\ref{sec:provenance} specifies the sample
provenance, dating, and track-retention criteria that fix the effective
exposure time $t_{\text{exp}}$ of a candidate mica slab.
Section~\ref{sec:dmflux} justifies the Standard Halo Model as a baseline
for the time-averaged dark matter flux over this exposure.
Section~\ref{sec:sensitivity} folds the overburden attenuation, the
melt-track and bore-channel thresholds, and the flux integral into projected
$90\%$ C.L.\ exclusion contours in the opaque and diffuse limits.
Section~\ref{sec:revisiting} then revisits the prior monopole-recast
dark matter scattering exclusions in light of these results.

\subsection{Mica Geology, Provenance, and Dating}
\label{sec:provenance}

\subsubsection{Sample Provenance and Selection Criteria}

Muscovite mica most suitable for macroscopic paleodetector applications will satisfy several criteria: sheets must be mechanically 
cleavable over areas of order $10^2$--$10^3~\text{cm}^2$ (for reasons of readout efficiency), free of inclusions and 
cleavage-plane discontinuities that would interrupt XRF mapping, and sourced from deposits with well-constrained thermal histories. Pegmatite-hosted muscovite satisfies these requirements more reliably than any other geological occurrence. 
Pegmatites crystallize from volatile-enriched, late-stage granitic melts under 
conditions that favor extreme grain growth; muscovite ``books'' with individual 
cleavage areas exceeding $10^3~\text{cm}^2$ are routinely recovered from 
commercial pegmatite operations \cite{cerny2005}. The layered silicate 
structure of muscovite, with its perfect \{001\} cleavage along the $ab$-plane, 
is a direct consequence of the $\text{KAl}_2(\text{AlSi}_3\text{O}_{10})
(\text{OH})_2$  silicate architecture: the basal plane across the mica ``sheets" is mechanically weak relative to intra-layer bonds, allowing controlled cleaving to a surface roughness on the order of tens of nanometers using a razor blade or adhesive tape.

Precambrian cratons provide the most favorable thermal environments 
for long-term track retention. The Grenville and Superior tectonic provinces of 
the Canadian Shield host large-book muscovite pegmatites with crystallization 
ages ranging from $\sim 1.0$ to $2.6~\text{Gyr}$, and have experienced 
minimal metamorphic or hydrothermal feature erasure (``overprinting'') since the Mesoproterozoic~\cite{rivers2008,easton2000}. For the purposes of this work, we treat sample selection and 
geochronological validation as a methodological program rather than reporting 
on specific samples; the criteria below define what a suitable paleodetector 
sample must satisfy before exposure limits can be computed.

\subsubsection{Track Retention and Thermal Closure}

The effective dark matter integration time $t_\text{exp}$ is not just governed by the crystallization ($i.e.$, formation) age of the sample, but also by its thermal history with respect to a track retention threshold often referred to as an ``annealing temperature.'' We expect that melt tracks produced by a transiting composite 
dark matter state will be features whose persistence over 
geological time is determined by the competition between the energy stored in 
the disordered lattice and the thermal activation energy for solid-state 
diffusion of displaced constituents back to equilibrium sites. This process 
is directly analogous to the annealing of fission tracks, which has been 
extensively characterized in muscovite.

Fission track annealing in muscovite proceeds through a distributed set of
thermally activated processes characterized by an Arrhenius-like dependence
of the annealing rate on temperature. Laboratory annealing experiments
establish a \emph{partial annealing zone} (PAZ) of approximately
$280$--$350^{\circ}\text{C}$ ($553$--$623~\text{K}$)~\cite{naeser1969,crowley1985}.
For temperatures sustained well below this range, the extrapolated annealing
timescale exceeds the age of the Solar System, so that on
$\mathcal{O}(\text{Gyr})$ timescales tracks are effectively permanent; for
temperatures within the PAZ, partial fading occurs on geological timescales;
above the PAZ, complete erasure occurs within the geological dwell time.

The \emph{closure temperature} $T_c$ for muscovite fission tracks, defined 
following \cite{1973CoMP...40..259D} as the temperature at which the system 
becomes closed to track loss during monotonic cooling, is approximately 
$320$--$350^\circ\text{C}$ ($\approx 600~\text{K}$) for cooling rates of 
order $1$--$10^\circ\text{C}$/Myr characteristic of stable cratons 
\cite{naeser1969, crowley1985}. For the purposes of 
this work, we adopt $T_c \approx 325^\circ\text{C}$ ($\approx 600~\text{K}$) 
as a conservative estimate for the lower bound of reliable track retention 
in muscovite. Because some cratonic regions sustain temperatures as low as $\sim 600$~K at $20$~km depth \cite{majorowicz2019thermal}, right at our adopted closure temperature, a mica sample preserving unannealed spontaneous fission tracks could not have spent significant time substantially deeper than $\sim 20$~km. We use this as the basis for the burial-depth assumption invoked in Section~\ref{sec:sensitivity}.

\subsubsection{Geochronological Validation}

A suitable paleodetector sample requires two independent age determinations 
that together constrain both the crystallization age and the track retention 
age. The primary crystallization age is most reliably established using 
$^{87}$Rb/$^{86}$Sr or U/Pb geochronology. Muscovite is a strongly Rb-bearing 
phase (typically $\sim 200$--$600~\text{ppm}$ Rb), so crystals from a single location 
will have resolvable isochrons \cite{Wang2022}. 

The functional track retention age can be established independently using 
\textit{in situ} $^{238}$U spontaneous fission track counting. Muscovite 
contains trace uranium at typical concentrations of $\sim 1$--$10~\text{ppm}$. 
The spontaneous fission decay constant for $^{238}$U is 
$\Gamma_f = 8.46 \times 10^{-17}~\text{yr}^{-1}$, meaning each $^{238}$U 
nucleus has this probability per year of undergoing spontaneous fission and 
producing a damage track \cite{Fleischer1975}. The areal density $A_s$ 
of surviving fission tracks intersecting a freshly cleaved and etched 
surface, measured in tracks per unit area, is given by
\begin{equation}
    A_s = \frac{1}{2}\, \Gamma_f\, n_U\, R_e\, t_\text{exp},
    \label{eq:fissiontrack}
\end{equation}
where $n_U$ is the volumetric number density of $^{238}$U atoms 
(in atoms~cm$^{-3}$), $R_e \approx 7~\mu\text{m}$ is the etchable 
length of a single fission track in muscovite \cite{Fleischer1975}, 
and $t_\text{exp}$ is the duration over which tracks have accumulated 
without thermal resetting. The factor of $1/2$ accounts for the geometric 
fact that only those fission tracks originating within one track-length of 
the cleavage surface, and directed toward it, will intersect and be 
counted on that surface. A measured $A_s$ consistent with the 
$^{87}$Rb/$^{86}$Sr or U/Pb crystallization age can confirm that tracks have been 
retained without significant annealing; a deficit in $A_s$ relative 
to the Rb-Sr or U-Pb prediction indicates a thermal resetting event and 
constrains $t_\text{exp}$ to be shorter than the crystallization age.

This dual approach provides a self-consistent internal calibration: the 
Rb-Sr age supplies the upper bound on $t_\text{exp}$, while the fission 
track density supplies a direct, thermally integrated lower bound. A mineral 
lattice that has preserved spontaneous $^{238}$U fission fragments at the 
density expected from its isotope age can be trusted to have retained the 
hydrodynamic damage features produced by larger dark matter composites, 
which are larger in physical dimension to fission tracks and presumably subject to 
less restrictive annealing kinetics. For the sensitivity projections in Section~\ref{sec:sensitivity}, we adopt 
a benchmark integration time of $t_\text{exp} = 10^9~\text{yr}$.

\subsection{Dark Matter Flux and Velocity Distribution Over Geological Time}
\label{sec:dmflux}

The sensitivity projections in this work require the time-integrated dark
matter flux through the mica sample over an exposure time
$t_{\text{exp}} \sim 1~\text{Gyr}$, established in Section~\ref{sec:provenance}.
The number of composite transits scales as
$N \propto (\rho_{\chi}/M_{D})\langle v_{D}\rangle\,A\,t_{\text{exp}}$, so the
relevant astrophysical input is the time-averaged flux
$\langle\rho_{\chi} v_{D}\rangle$ integrated over the full geological
retention time. Because this average is taken over $\gtrsim 10^{9}$ orbital
periods of the Solar System about the Galactic center, the natural concern
is whether departures from a smooth, stationary halo introduce corrections
the SHM cannot capture. Below we argue that, for the time-averaged flux
relevant to paleo-detection, these effects are subdominant and that the SHM
provides a conservative baseline.

Three classes of departure from the SHM are worth assessing: (i)~stochastic
density enhancements from dark matter subhalos, (ii)~kinematic substructure
from past mergers, and (iii)~differences between the canonical Maxwellian
and velocity distributions extracted from hydrodynamical simulations of
Milky-Way analogues. Ref.~\cite{Ibarra:2019jac} finds that the probability
of the Earth presently residing inside a subhalo large enough to produce an
$\mathcal{O}(1)$ enhancement of the local density is
$\lesssim 7\times 10^{-4}$, with the dominant contribution to the local
$\rho_{\chi}$ unaffected. The dominant kinematic substructure in the local
neighborhood, debris from the Gaia Sausage Enceladus merger, was accreted
at redshift $z \sim 1$--$2$ and is largely phase-mixed by the present
epoch~\cite{Necib:2018iwb}; our sensitivity projection averages over all
incidence angles and so is insensitive to the residual radial anisotropy.
Finally, Ref.~\cite{Bozorgnia:2016ogo} extracts local velocity distributions
from the EAGLE and APOSTLE simulations and finds that Maxwellian fits with
peak speeds in the range $223$--$289~\text{km/s}$ reproduce direct-detection
event rates derived from the simulated distributions. One known and
potentially relevant deviation is the LMC-induced high-velocity tail
enhancement of Ref.~\cite{Bozorgnia:2025lmc}, which has been shown to
extend exclusion contours for plastic etch detectors at the highest masses.
Since this effect only adds high-velocity flux, neglecting it yields a
conservative bound for our purposes; we defer a quantitative treatment to
future work.

\subsection{Sensitivity Projections}
\label{sec:sensitivity}

The energy-deposition framework of Section~\ref{sec:models}, the
$R_{\min} = 25~\mu\text{m}$ readout calibration of
Section~\ref{sec:readout}, the geochronological prescription for
$t_{\text{exp}} \sim 10^{9}~\text{yr}$ of Section~\ref{sec:provenance},
and the astrophysical input of Section~\ref{sec:dmflux} are now in hand.
We will combine them in three steps. First, we follow a dark matter composite from
its astrophysical incidence velocity through the Earth overburden, computing
the energy and velocity with which it arrives at the mica slab. Second, we
determine the minimum residual velocity required to leave a detectable
signature, either as a melt track (Section~\ref{sec:models}) or as a
bore-channel feature (introduced shortly). Third, we integrate the composite
flux satisfying both conditions over a $1~\text{m}^{2}$ mica area and a
$10^{9}~\text{yr}$ exposure, yielding the projected $90\%$ exclusion
contours of Fig.~\ref{fig:bounds}. The melt-track signature is
unique in mica: no Standard Model process is known to produce
straight-line damage features extending continuously across a centimeter-scale
cleaved sheet.

\subsubsection{Overburden Attenuation}
We have already discussed signatures from energy deposition of dark matter into mica due to nuclear recoils in Section~\ref{sec:models}. However, as dark matter traverses the Earth en route to the mica sheet,
its energy is degraded by nuclear recoils, and we must account for this
attenuation to determine the kinematic state with which it arrives at the mica.

Dark matter passing through an overburden medium $M$ with density $\rho_{M}$ loses energy to nuclear recoils at a rate proportional to its kinetic energy. In the limit of dark matter composite masses much larger than nuclear masses, this rate takes simple forms for the two models of DM--nuclear interaction considered in this work. In the opaque limit where the composite has a potential large enough to repel all incoming atoms the energy loss is \cite{Acevedo:2020avd,Acevedo:2021kly} ($cf.$ Eq. \ref{eq:Edep_opaque})
\begin{equation}\label{eq:opaqueEloss}
    \frac{dE_D}{dx} = -\frac{2}{m_D}\rho_{M} \sigma_{\text{geo}} E_D.
\end{equation}
In the diffuse/loosely-bound limit, we can model the energy deposition of nuclei hitting dark matter constituents \cite{Acevedo:2024lyr} inside the DM composites as, ($cf.$ Eq. \ref{eq:Edep_diffuse_macro})
\begin{equation}\label{eq:diffuseEloss}
    \frac{dE_D}{dx} = -\frac{2}{m_D}\left(\sum_{A \in M}A^4 f_{A}\right) \rho_{M} \sigma_{n} E_D,
\end{equation}
where we sum over all nuclides in the medium $M$, labelled by mass number $A$ and with mass abundance $f_{A}$. To bring Equations~\ref{eq:opaqueEloss} and~\ref{eq:diffuseEloss} into a common form, we define a medium-dependent \textit{nuclide moment} $\kappa_M$,
\begin{equation}\label{eq:momentDfn}
\kappa_M =
\begin{cases}
  1 & \text{in the opaque case},  \\
  \displaystyle\sum_{A \in M}A^4 f_{A} & \text{in the diffuse case,}
\end{cases}
\end{equation}
and denote either $\sigma_{\text{geo}}$ or $\sigma_{n}$ generically as $\sigma$. Equations~\ref{eq:opaqueEloss} and~\ref{eq:diffuseEloss} then take the unified form
\begin{equation}\label{eq:genericEloss}
    \frac{dE_D}{dx} = -\frac{2}{m_D}\kappa_M\rho_{M} \sigma\, E_D.
\end{equation}
This kinematic energy loss arises from elastic nuclear scattering: each target nucleus of mass $m_N$ receives a recoil energy $E_R \approx m_N v_D^2$ in the heavy-composite limit, independent of the lattice environment. The composite loses this energy regardless of whether the recoiling atom is displaced from its lattice site or vibrates back and deposits $E_R$ as phonons.

Solving the differential equation in Equation \ref{eq:genericEloss} gives
\begin{equation}\label{eq:energyExp}
    E_f = E_0e^{-2\frac{\sigma}{m_D}\mathcal{I}},
\end{equation}
where we define the integral $\mathcal{I}$ over the dark matter's path through the overburden,
\begin{equation}\label{eq:overburden}
   \mathcal{I} = \int_{\substack{\mathrm{path}}} \kappa_{\mathrm{Earth}}(x)\,\rho_{\mathrm{Earth}}(x)\, dx.
\end{equation}
Equivalently, the speed of dark matter after traversing the overburden is,
\begin{equation}\label{eq:velocityExp}
    v_f = v_0e^{-\frac{\sigma}{m_D}\mathcal{I}}.
\end{equation}
We assume that dark matter travels in a straight line through the Earth, and that the resulting collisions with nuclei do not significantly deflect that path.
 To see why, recast Equation~\ref{eq:energyExp} in the form $E_f = E_0e^{-2\frac{m_n}{m_D}N}$, where $m_n$ is the characteristic mass of nuclei in the Earth ($\sim 30$ GeV) and $N$ is the number of DM-nuclear collisions throughout the entire overburden. From this it is clear that the dark matter will lose almost all of its energy and will fail to melt a track once $N \gtrsim m_D / m_n$. On the other hand, basic kinematic considerations show that the angle of deflection from a single collision goes as $\delta \theta\sim \frac{m_n}{m_D}$. But, the deflection is equally likely to be in any direction transverse to the dark matter's velocity, meaning the deflection will on average be zero. However, there will still be deflection over time, as one should treat the deflections as a random walk, with a typical deviation after $N$ collisions of $\Delta \theta\sim \frac{m_n}{m_D}\sqrt{N}\sim \sqrt{\frac{m_n}{m_D}}$. But, even for dark matter as light as $m_D = 10^{10}$ GeV, the angle of deflection over the entire overburden is $\Delta\theta \sim 10^{-4}$.

In contrast, gravity will affect the path of dark matter once the overburden slows it down enough. Gravity becomes relevant only once the overburden has slowed the composite
to roughly the Earth's surface escape speed,
$v_{\mathrm{grav}} = \sqrt{G M_{\mathrm{Earth}}/R_{\mathrm{Earth}}} \approx 7.9$~km/s.
As we will see below, this regime is reached only in a corner of the
parameter space at simultaneously high mass and high cross-section. Properly
modelling the resulting curved trajectories and possible Earth capture is
beyond the scope of this work; we note, however, that neglecting them yields
a conservative projection. Including gravity could only affect the bound in
the high-mass, high-cross-section corner of parameter space, which from
Figure~\ref{fig:bounds} is a small fraction of the projected exclusion
region.

We consider all dark matter trajectories through the Earth that hit the mica, including those going through the Earth's mantle and core. For density profile of the Earth, we use the preliminary Earth reference model (PREM) \cite{Dziewonski:1981xy}, which expresses the density of the Earth as a function of depth. For $\kappa_{\mathrm{Earth}}$ in the diffuse case, we need the elemental abundance as a function of depth, which we take from \cite{Morgan6973,MCDONOUGH2003547,clarke1924composition,WANG2018460} and record in Table \ref{tab:earthElemComp}. Conveniently, PREM reports the densities as a piecewise quartic polynomial. This allows the integral in Equation \ref{eq:overburden} to be computed analytically for a straight-line trajectory through the Earth, which significantly speeds up the computation.

\begin{figure}[t]
\centering
\includegraphics[width=0.46\textwidth]{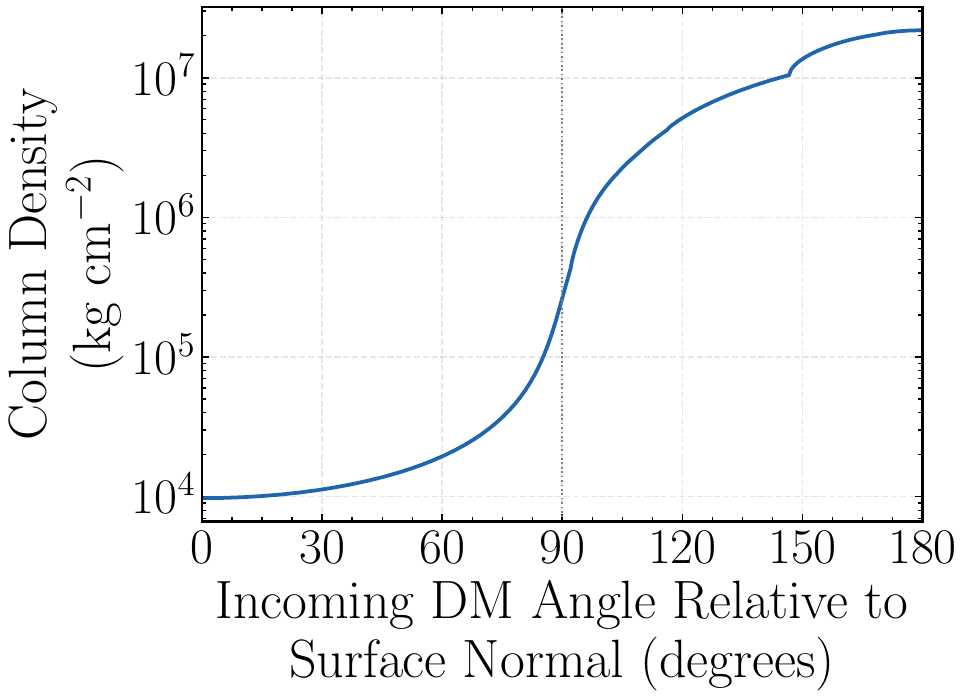}
\hspace*{0.1 cm}
\includegraphics[width=0.48\textwidth]{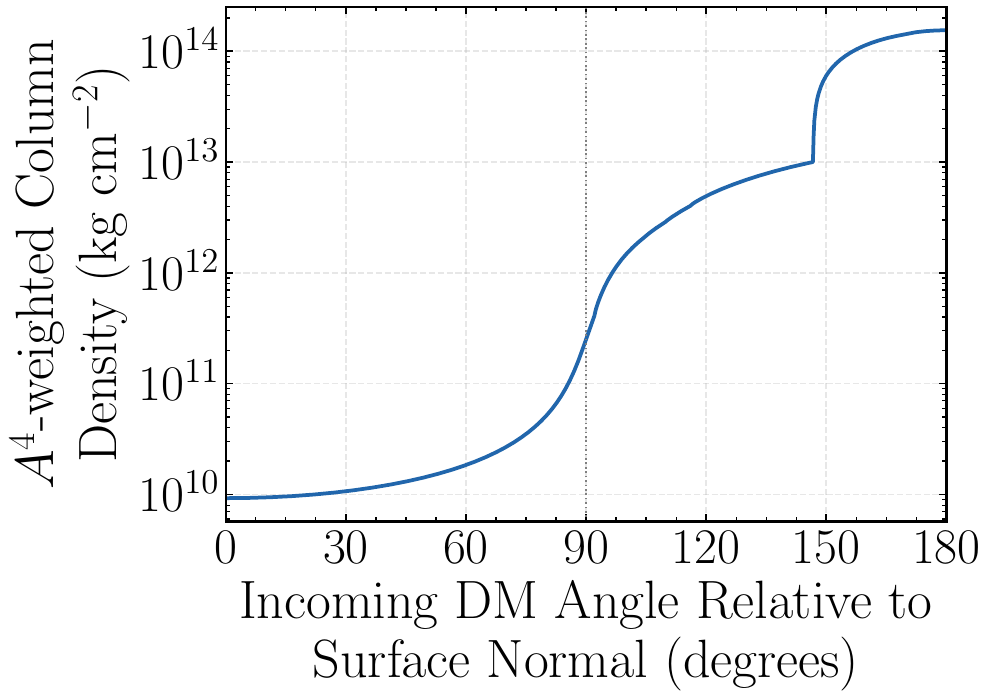}
\caption{The integral defined in Equation \ref{eq:overburden} for a straight-line path through the overburden hitting the mica at some angle to the Earth's surface normal for the opaque (left) and diffuse (right) cases. In the opaque case, the integral gives the column density of the overburden along the dark matter's path and in the diffuse case, it gives the sum weighted by $A^4$ over all nuclides in the overburden.}\label{fig:overburden}
\end{figure}

The integral over the overburden for a straight line path is plotted in Figure \ref{fig:overburden} for both the opaque and diffuse case. It is plotted vs the path's angle relative to the Earth's surface normal so that, at $0\degree$, the dark matter falls vertically through the crust and at $180\degree$, the dark matter travels vertically up the Earth's core. For the opaque case, this integral is simply the column density through that path. For the diffuse case, it is the column density of each nuclide, weighted by $A^4$.

\begin{table}[h]
\centering
\begin{tabular}{lccccccccc
cccc}
\hline
 & $^{16}$O & $^{28}$Si & $^{27}$Al & $^{56}$Fe & $^{40}$Ca & $^{23}$Na & $^{39}$K & $^{24}$Mg & $^{48}$Ti & $^{57}$Ni & $^{59}$Co & $^{31}$P & $^{32}$S \\
\hline
Crust & 46.7 & 27.7 & 8.1 & 5.1 & 3.7 & 2.8 & 2.6 & 2.1 & 0.6 & -- & -- & -- & -- \\
\hline
Mantle & 44.3 & 21.3 & 2.3 & 6.3 & 2.5 & -- & -- & 22.3 & -- & 0.2 & -- & -- & -- \\
\hline
Core & -- & -- & -- & 84.5 & -- & -- & -- & -- & -- & 5.6 & 0.3 & 0.6 & 9.0 \\
\hline
\end{tabular}
\caption{Mass percentages of the most abundant elements in the Earth's
crust, mantle, and core. Adopted in Eq.~\ref{eq:overburden} when evaluating
the diffuse-case nuclide moment $\kappa_{\mathrm{Earth}}$ along trajectories
that traverse each layer.}\label{tab:earthElemComp}
\end{table}

\subsubsection{Melt threshold}
We adopt $R_{\mathrm{min}} = 25~\mu\text{m}$ as the benchmark minimum detectable defect radius established by the XRF calibration in Section~\ref{sec:readout}. The defect threshold defines a critical entry velocity below which the composite cannot produce a melt track of radius $R_{\mathrm{melt}} \geq R_{\min}= 25~\mu\text{m}$. 

In the opaque case, from the adiabatic melt scaling (Eq.~\ref{eq:Rmelt_adiabatic}), $R_{\mathrm{melt}} = R_D \sqrt{\eta v_D^2 / (C_p \Delta T_{\mathrm{melt}} + H_f)}$, and the melt track shrinks below $R_{\min}$ when
\begin{equation}\label{eq:vmeltOpaque}
    v_D < v_{\mathrm{melt}} \equiv \frac{R_{\min}}{R_D}\sqrt{\frac{C_p \Delta T_{\mathrm{melt}} + H_f}{\eta}} \approx 1.4~\text{km/s}\times \frac{R_{\min}}{R_D}.
\end{equation}
However, for opaque composites $R_D>R_{\mathrm{min}}$, even if the composite deposits enough total energy to melt a hole of radius $R_{\mathrm{melt}}$, that energy will be spread over an area of radius $R_D>R_{\mathrm{melt}}$ and may not produce melting. Instead, the dark matter will bore through the mica. We treat this in Section \ref{sec:bore}.

In the diffuse case, the melt velocity is determined by two free parameters in addition to the per-nucleon cross section $\sigma_n$: the dark matter composite's physical size $R_D$, and the mass-averaged optical depth for a nucleus in mica, $\tau_{\mathrm{mica}}$. Fortunately, these three quantities can be related by mass-averaging Equation \ref{eq:opticalDepth2} to get,
\begin{equation}\label{eq:sizeVsOptDepth}
    \tau_{\mathrm{mica}} = \frac{\kappa_{\mathrm{mica}}\sigma_n}{\pi R_D^2},
\end{equation}
where $\kappa_\mathrm{mica}$ is mica's nuclide moment, defined in Equation \ref{eq:momentDfn}. With the chemical formula for mica, $\text{KAl}_2(\text{AlSi}_3\text{O}_{10})(\text{OH})_2$, we find $\kappa_\mathrm{mica} \approx 5.023\times 10^{5}$ and,
\begin{equation}\label{eq:sizeVSdiffuseCrossSection}
    R_D \approx 400\sqrt{\frac{\sigma_n}{\tau_{\mathrm{mica}}}}.
\end{equation}
With a fixed choice of optical depth, we can relate the per-nucleon cross section and the physical size of the composite. Like the opaque case, we have two regimes determined by $R_D$. If $R_D$ is less than our detection threshold $R_{\mathrm{min}}=25~\mu\text{m}$ then the composite must be in high-melt regime to melt a detectable track. Using Equation \ref{eq:highMeltRadius}, the melt track shrinks below $R_{\min}$ when,
\begin{equation}\label{eq:vmeltHigh}
    v_D < v_{\mathrm{melt}} \equiv \frac{R_{\min}}{R_D}\sqrt{\frac{C_p \Delta T_{\mathrm{melt}} + H_f}{\eta \tau_{\mathrm{mica}}}} \approx 3.6~\text{m/s}\times \frac{R_{\min}}{\sqrt{\sigma_n}}.
\end{equation}
If $R_D > R_{\mathrm{min}}=25~\mu\text{m}$ then the composite could be in the low-melt regime and still melt a detectable track. Using Equation \ref{eq:lowMeltRadius} instead, the melt track shrinks below $R_{\min}$ when,
\begin{equation}\label{eq:vmeltLow}
    v_D < v_{\mathrm{melt}} \equiv \frac{\frac{R_{\min}}{R_D}\sqrt{\frac{C_p \Delta T_{\mathrm{melt}} + H_f}{\eta \tau_{\mathrm{mica}}}}}{\sqrt{1-\left(1-\left(R_{\min}/R_D\right)^2\right)^{\frac{3}{2}}}} \approx \frac{3.6~\text{m/s}\times R_{\min}/\sqrt{\sigma_n}}{\sqrt{1-\left(1-\left(R_{\min}/R_D\right)^2\right)^{\frac{3}{2}}}}.
\end{equation}

\subsubsection{Sub-melt detection:  the boring regime}\label{sec:bore}

For opaque composites with $R_D \geq R_{\min}=25~\mu\text{m}$, there exists a detection channel beyond the melt-track regime. Even after the composite has decelerated below $v_{\mathrm{melt}}$ and can no longer produce a surrounding melt region, it continues to physically bore a cylindrical puncture of radius $R_D$ through the mineral lattice. This puncture is a region where every atom within the geometric cross-section has been mechanically displaced by the transiting composite, producing a hollow channel that constitutes a permanent structural defect in the crystal. Such a feature is detectable by XRF transmission mapping provided $R_D > R_{\min}$, since the void exposes the copper backing over the full composite cross-section, producing a Cu~K$\alpha$ enhancement.

To produce a contiguous bore with radius $R_{\mathrm{bore}}$, the dark matter must deposit energy per unit path length exceeding,
\begin{equation}\label{eq:energyThresholdBore}
    \left|\frac{dE_D}{dx}\right|_{\mathrm{bore}} = \rho  H_f \pi R_{\mathrm{bore}}^2,
\end{equation}
where $H_f$ is the energy needed to displace all nuclei in a volume from the lattice, which we take to be equal to the specific latent heat of fusion. Equating this with the energy deposition by an opaque composite traversing the medium, we find a hard velocity cutoff, below which the composite cannot produce a bore hole regardless of its size,
\begin{equation}\label{eq:vbore}
    v_D < v_{\mathrm{bore}} \equiv \sqrt{H_f} \approx 0.63~\text{km/s},
\end{equation}
where we have used $H_f \approx 4 \times 10^5$~J/kg. For opaque composites $R_D \geq R_{\min}=25~\mu\text{m}$, we use this velocity threshold over Equation \ref{eq:vmeltOpaque}.

The combination of the two thresholds opens a kinematic window
$v_{\mathrm{bore}} < v_D < v_{\mathrm{melt}}$ for opaque composites with
$R_D \geq R_{\min}$. A composite arriving at the mica with a velocity in
this window, typical of heavy composites whose initial halo velocity has
been substantially attenuated by overburden, produces a sub-melt void of
radius $R_D$ rather than a surrounding melt halo. Because such a void
exposes the copper backing over the full geometric cross-section, it is
detectable by the XRF transmission method whenever $R_D > R_{\min}$; for
$R_D < R_{\min}$ the void falls below the imaging resolution and this
channel does not contribute. The void signature is qualitatively distinct
from a melt track: it lacks the vitrified halo of resolidified material
that surrounds a melt feature, and instead presents as a clean cylindrical
absence of lattice material, so that follow-up optical microscopy of any
candidate XRF feature could in principle distinguish the two.

\subsubsection{The projected sensitivity region}

To produce a detectable track, the incoming dark matter must retain sufficient energy after traversing the overburden and the mica slab itself. Since we assume the dark matter travels in a straight line, the path through the overburden and therefore the integral in Equation \ref{eq:overburden} depends only on the dark matter's initial incoming direction: $\mathcal{I}=\mathcal{I}(\hat{v}_D)$. 

For a given dark matter mass $m_D$ and cross section $\sigma$, dark matter at incoming astronomical velocity $\mathbf{v}_D=v_D \hat{v}_D$ will make a track if,

\begin{equation}
\label{eq:vminCondition}
    v_D > v_{\mathrm{min}}(\hat{v}_D)= v_{\mathrm{track}}e^{\frac{\sigma}{m_D}\mathcal{I}(\hat{v}_D)},
\end{equation}
where, in the opaque case $v_{\mathrm{track}}$ is given by Equation \ref{eq:vbore} if $\sqrt{\frac{\sigma_{\mathrm{geo}}}{\pi}}=R_D>25 \:\mu$m and by Equation \ref{eq:vmeltOpaque} if $R_D<25 \:\mu$m. In the diffuse case, we fix a choice of optical depth.  $v_{\mathrm{track}}$ is then given by Equation \ref{eq:vmeltHigh} if $400\sqrt{\frac{\sigma_n}{\tau_{\mathrm{mica}}}}=R_D>25 \:\mu$m and by Equation \ref{eq:vmeltLow} if $R_D<25 \:\mu$m.

Assuming an analysis observed no melt tracks or bore holes, and taking Poisson statistics, a 90\% confidence upper limit would correspond to an expected number of events $N_{\mathrm{obs}} = -\ln(0.1) \approx 2.3$. The expected number of detectable dark matter transits through a mica slab of area $A$ over exposure time $t_{\mathrm{exp}}$ is
\begin{align}\label{eq:fullNumDMTime}
    \begin{split}
        N_{\mathrm{exp}}(m_D,\sigma) = \frac{A\,\rho_{\mathrm{DM}}}{m_D}&\int_0^{t_{\mathrm{exp}}} dt\int_{v_{\mathrm{min}}(\hat{v}_D,t)}^\infty dv_D\; v_D^2 \int d\hat{v}_D\; f(-\mathbf{v}_D\mid t)\;\mathbf{v}_D\cdot\hat{n}(t)\;E\!\left(\arccos(\hat{v}_D \cdot\hat{n}(t))\right)
    \end{split}
\end{align}
where $E(\alpha)$ is the probability of dark matter making a track when it hits at an angle of $\alpha$ to the mica's surface normal, $f_{\mathrm{vel}}(\mathbf{v}\mid t)$ is the local dark matter velocity distribution, and $\hat{n}(t)$ is the mica surface normal.

This formula as written allows for the dark matter velocity distribution, the depth and orientation of the mica with respect to the Earth, and the orientation of the Earth's surface normal above the mica with respect to galactic coordinates to vary over time. However, we make several simplifying assumptions. As discussed in Section \ref{sec:dmflux}, the first is that the dark matter velocity distribution is unchanging over the $10^9$ yr exposure time and that it follows the Standard Halo model with modern parameters,
\begin{equation}\label{eq:shm}
    f_{\mathrm{vel}}(\mathbf{v}) \propto \exp\left(-\frac{\|\mathbf{v}-\mathbf{v}_s\|^2}{v_\chi^2}\right)\Theta(v_{\rm esc}-\|\mathbf{v}-\mathbf{v}_s\|),
\end{equation}
where we take the escape speed of the MW to be $v_{\rm esc} = 503\ \text{km/s}$\cite{Deason2019}, the circular velocity at the Sun's position to be $v_\chi=222 \: \text{km/s}$\cite{Eilers_2019dmspeed} and the Sun's speed relative to the galaxy to be $v_s=232~\text{km/s}$~\cite{Schoenrich:2009bx}.
The second assumption relates the orientation of the Earth's normal above the mica with respect to the galaxy. To calculate this exactly over $10^9$~yr, one would have to fold in the Earth's daily rotation, axial precession, the orbital motion of the Sun about the Galactic center, the Sun's oscillation through the Galactic plane, and the tectonic drift of the mica sample's host pegmatite in latitude. We will assume that these myriad effects average out and the Earth's surface normal is randomly distributed over all directions over the billion-year exposure time. This allows us to treat the incoming dark matter as effectively isotropic, reducing the dark matter \textit{velocity} distribution $f_{\mathrm{vel}}(\mathbf{v})$ to the dark matter \textit{speed} distribution, 
\begin{equation}\label{eq:speedDist}
    f_{\mathrm{speed}}(v_D) = \int d\hat{v}_D\; f_{\mathrm{vel}}(\mathbf{v}_D).
\end{equation}
The third simplifying assumption is that the probability of making a track is $100\%$ if the dark matter hits within some acceptance half-angle $\theta_{\mathrm{acc}}$ and $0\%$ otherwise, meaning $E(\alpha) = \Theta(\theta_{\mathrm{acc}}-\alpha)$, where $\Theta$ is the Heaviside step function. We set $\theta_{\mathrm{acc}} = 60 \degree$.
Next, even though we treat the astrophysical dark matter as isotropic, the dark matter will not be isotropic by the time it reaches the mica because of the significant dependence of the overburden on the direction of incoming dark matter (see Figure \ref{fig:overburden}). Depending on the incoming direction, the dark matter may or may not have enough energy to leave a detectable signature. Therefore Equation \ref{eq:fullNumDMTime}, and specifically $v_{\mathrm{min}}$, depends on the mica's depth and orientation. Finally, we conservatively fix the burial depth of the mica at $20$~km
throughout the exposure: below this depth the geothermal temperature
exceeds the muscovite fission-track closure temperature (Section~\ref{sec:provenance}),
so any sample with observable spontaneous fission tracks can be assumed
not to have spent significant time deeper. In the full analysis, this will have to be further investigated for particular mica samples, given models for the dynamical evolution of the pegmatite formation encasing any given mica sample. 

Under these assumptions, the number of expected tracks for a given $m$ and $\sigma$ is,
\begin{align}
\label{eq:fullNumDMSimple}
    \begin{split}
        N_{\mathrm{exp}}(m_D,\sigma) = \frac{A\,\rho_{\mathrm{DM}}\,t_{\mathrm{exp}}}{4\pi m_D}& \int d\hat{v}_D\;\;\hat{v}_D\cdot\hat{n}\;\Theta(\mathbf{v}\cdot \hat{n}-\cos(\theta_{\mathrm{acc}}))\left(\int_{v_{\mathrm{min}}(\hat{v}_D)}^\infty dv_D\; v_D^3 \,f_{\mathrm{speed}}(v_D)\right).
    \end{split}
\end{align}
Here, $\Theta$ is the Heaviside step function. The outer integral is over the unit sphere and the factor of $4\pi$ ensures the isotropic velocity distribution is properly normalized. This integral must be done numerically. On the other hand, by assuming dark matter follows the Standard Halo model, the inner integral is analytically computable for any given $v_{\mathrm{min}}$, which greatly speeds up computation. The sensitivity region is then the set of $(m_D, \sigma)$ for which $N_{\mathrm{exp}} \geq 2.3$. This was found with numerical root finding.

\begin{figure}[t]
\centering
\includegraphics[width=0.48\textwidth]{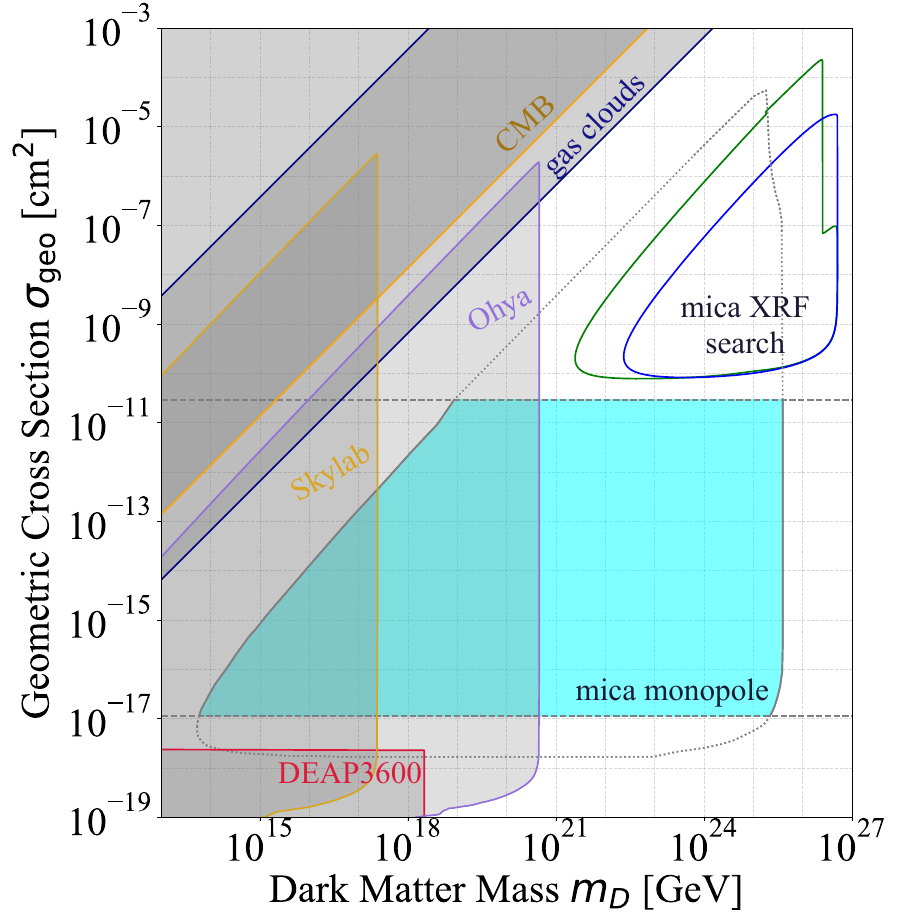}
\hspace{0.1 cm}
\includegraphics[width=0.48\textwidth]{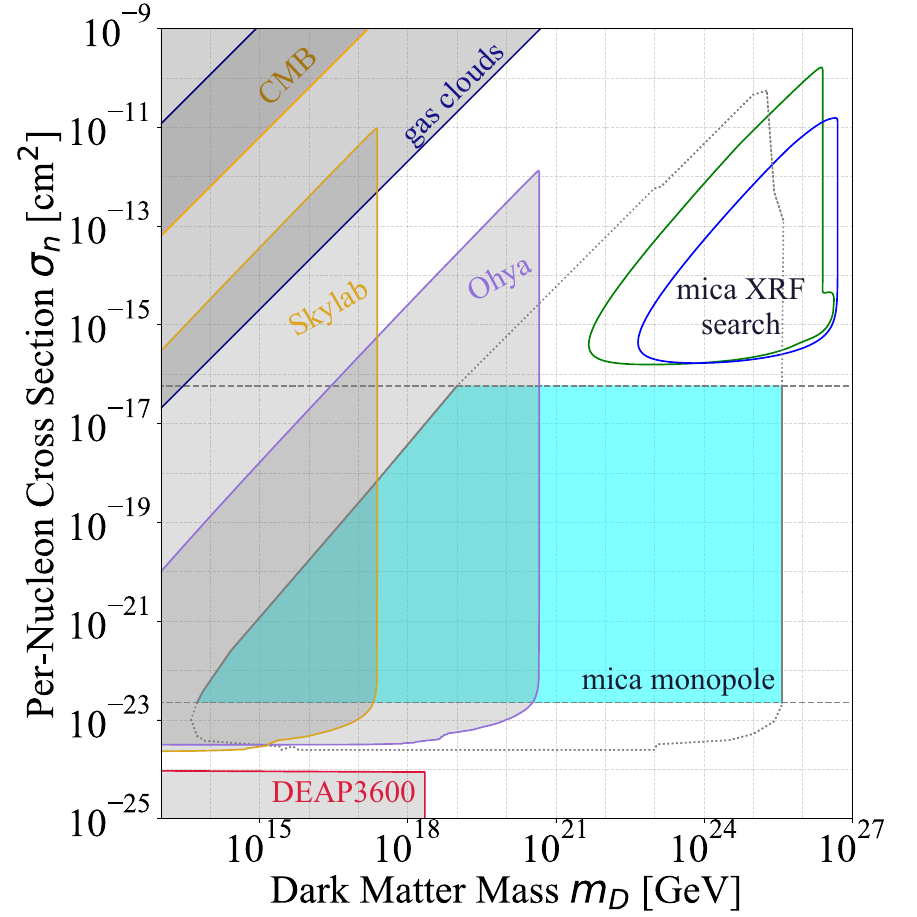}
    \caption{Projected $90\%$ C.L.\ sensitivity to heavy composite dark matter
from a benchmark exposure of $1~\text{m}^{2}\times 10^{9}~\text{yr}$ of cratonic muscovite, for opaque-geometric (left) and diffuse loosely-bound
(right) interaction regimes of Section~\ref{sec:models}. The diffuse-case
projection fixes the mass-averaged optical depth in mica to
$\tau_{\mathrm{mica}}=0.1$. Solid green and blue curves bracket the two
extremal cleavage plane orientations relative to the Earth's surface:
parallel to it (green) and perpendicular to it (blue), and the true
sensitivity for any given sample lies between them. Existing constraints
shown for comparison: cosmic microwave background \cite{Gluscevic:2017ywp},
Galactic Center gas clouds \cite{Bhoonah:2020dzs}, DEAP-3600
\cite{DEAPCollaboration:2021raj}, and the recast plastic-etch nuclearite
searches on Skylab and in the Ohya mine
\cite{Bhoonah:2020fys,Bozorgnia:2025lsl}. The dotted curve is the recast of
the Price--Salamon monopole mica search of
Refs.~\cite{Acevedo:2021tbl,Jacobs:2014yca}; as argued in Section~\ref{sec:revisiting}, only the blue-shaded band inside this contour
is recastable as a robust dark matter bound. The blue-shaded band is bounded below by
the fission-track retention threshold (alpha-recoil tracks anneal on
$\sim$\,Myr timescales and so are unreliable proxies for Gyr-old composite
damage) and above by the threshold at which the inferred melt features
would have caused the original samples to be rejected during the
``optically clean'' visual-selection cut.}
\label{fig:bounds}
\end{figure}

We present our projected bounds for the opaque and diffuse cases in Figure \ref{fig:bounds}, assuming a local dark matter density of $0.3~\text{GeV}/\text{cm}^3$, a $1~\text{m}^{2}$ mica area
and a $10^{9}~\text{yr}$ exposure. For the diffuse case, we set $\tau_{\mathrm{mica}}=0.1$. We compute the bounds in the two extremal cases, when the mica's surface is parallel to with the Earth's surface and when the mica surface is perpendicular to the Earth's surface, to capture how the sensitivity region may depend on the mica's changing orientation in the Earth over geological time. In Section~\ref{sec:revisiting}, we argued that the robust recastable parameter space from the Price--Salamon and Snowden-Ifft searches lies between the fission-track damage threshold and melt features at $R_{\mathrm{melt}} \gtrsim 10~\mu\text{m}$. In Figure \ref{fig:bounds}, we also overlay the earlier recast bounds from Refs.~\cite{Acevedo:2021tbl, Jacobs:2014yca} as dashed contours, and demarcate the melt-track projections of this work (solid contours) at benchmark melt radii of $R_{\mathrm{melt}} = 25$ and $100~\mu\text{m}$. We also compare our result with existing bounds from the cosmic microwave background (CMB) \cite{Gluscevic:2017ywp}, interstellar gas clouds \cite{Bhoonah:2020dzs,Bhoonah:2018gjb,Bhoonah:2018wmw}, DEAP3600 \cite{DEAPCollaboration:2021raj}, and exclusions from plastic etch detectors aboard the Skylab satellite as well as the Ohya mine \cite{Bhoonah:2020fys,Bozorgnia:2025lsl} (see also ~\cite{Acevedo:2020gro, Ray:2023auh,Bhattacharya:2024pmp}).

\subsection{Revisiting dark matter searches using mica monopole data}
\label{sec:revisiting}

The strongest existing constraints on heavy composite dark matter from mineral detectors~\cite{Acevedo:2021tbl,Jacobs:2014yca} arise from recasts of the Price–Salamon monopole search~\cite{Price:1986ky}. In this search, $\sim\!1200$~cm$^{2}$ of Precambrian muscovite were scanned to look for anomalous etchable tracks, with no candidates observed, satisfying a sextuple-coincidence criterion at nuclear stopping cross-section $S_{n} > 2.42$~GeV\,cm$^{2}$\,g$^{-1}$. Snowden-Ifft
\textit{et al.}~\cite{SnowdenIfft:1995ke} later searched
$80{,}720~\mu\text{m}^{2}$ of 0.5~Gyr muscovite for WIMP-recoil etch pits using atomic force microscopy, also obtaining null results. The resulting limits on heavy composite dark matter with large nuclear scattering cross sections~\cite{Acevedo:2021tbl,Jacobs:2014yca}, which essentially rely on recasts of the Price–Salamon search, are subject to few shortcomings (which we will discuss below) that affect their robustness. In particular, these analyses rest on two key assumptions: (i) that alpha-induced track damage provides a reliable proxy for dark matter–induced tracks, and (ii) that such tracks are not significantly erased by thermal annealing over geological timescales.\\

A direct comparison with the present work requires a brief recap of the
original Price-Salamon methodology. Their analysis used ten muscovite
samples of total area $\sim 1200~\text{cm}^{2}$, selected as ``optically
clean and free of mechanical defects'' before any chemical processing.
Following selection, each sample was cleaved into six co-aligned sheets,
etched in $48\%$ HF, and inspected by optical microscopy; a candidate event
was required to register an etched feature in coincident positions on
all six adjacent sheets, a sextuple-coincidence criterion designed to
suppress shallow alpha-recoil pits and chemical etch artifacts. The
threshold for an alpha-recoil-like etched feature corresponded to a nuclear
stopping cross-section of $S_n > 2.42~\text{GeV}\,\text{cm}^{2}\,\text{g}^{-1}$,
adopted by Refs.~\cite{Acevedo:2021tbl,Jacobs:2014yca} as the effective
damage threshold for their dark matter recast. The non-destructive
XRF readout developed in this work makes neither the optically-clean
selection cut nor the assumption of alpha-track retention; it instead
images the post-cleave mica directly and depends only on the much weaker
condition that melt-track or bore-channel damage be retained at the
fission-track-retention temperature, which is independently verified by
the spontaneous-track count of Section~\ref{sec:provenance}.

Recent thermochronological studies cast significant doubt on the validity of the second assumption.  Yuan
\textit{et al.}~\cite{yuan2009annealing} measured the annealing kinetics of alpha-recoil tracks in phlogopite mica and found closure temperatures of only $26$--$37\,^{\circ}$C for geologically relevant cooling rates ($1$ -$100\,^{\circ}$C\,Myr$^{-1}$), implying a complete annealing on $\sim\!10^{6}$~yr timescales at temperatures well below typical upper crust temperatures.  To our knowledge, no equivalent annealing study of alpha-recoil tracks in muscovite has been performed. Despite absence of such data, and given that alpha-recoil closure temperatures in phlogopite fall hundreds of degrees below the well-established fission-track closure temperature in muscovite ($\gtrsim 300\,^{\circ}$C) \cite{Fleischer1975,Wagner1992}, the assumption that alpha-level damage persists in muscovite (which is very similar to phlogopite) over Gyr timescales is not
empirically supported.\\
In addition, Price and Salamon restricted their analysis to optically clean, undamaged mica sheets~\cite{Price:1986ky}. As we show in this work, in the large-interaction regime heavy composite dark matter would produce melt tracks, leaving visible melt inclusions in the mineral lattice. Such samples would likely have been rejected during the initial visual selection stage, which explicitly favored clean, unblemished specimens.  Refs.~\cite{Acevedo:2021tbl,Jacobs:2014yca} overlooked this sample-selection cut and consequently derived exclusions extending to much larger interaction cross-sections. As a result, those limits are not reliable in the high–scattering cross-section regime. In Section~\ref{sec:sensitivity}, we identify the region of parameter space where the recast remains trustworthy and explicitly delineate this band, enabling a consistent comparison with the analysis presented in this work. The melt-track search proposed in this work targets the high-damage portion of this region using an independent, non-etch-based readout. 
\section{Conclusion}
\label{sec:conclusion}
In this work we have developed a quantitative framework, an experimental
readout demonstration, and geochronological calibration for muscovite mica as a paleodetector sensitive to heavy composite dark matter,
in the regime of composites with radii ranging from nanometers to microns,
where existing etching-based readout methods are inefficient.

We modeled energy deposition by transiting composites in both an opaque
(geometric) limit and a diffuse (loosely-bound) limit using a Sedov-Taylor
thermal spike formalism, deriving analytic scalings for the melt track radius
as a function of composite radius in each case. SRIM/TRIM simulations of 
nuclear recoil cascades calibrated the phonon efficiency factor governing local energy 
deposition to $\eta \simeq 0.75$, and reproduced the analytic melt scalings at
sub-micron composite radii, where the cascade picture remains tractable. 
We also identified a detection mode for large opaque composites 
that have been substantially decelerated by overburden but retain sufficient kinetic 
energy to bore a contiguous puncture through the crystal.

We demonstrated a non-destructive X-ray fluorescence transmission readout
using a copper-backing contrast technique, capable of identifying
micron-scale damage features in cleaved mica sheets over macroscopic scan
areas. Calibration against laser-ablated proxy features established a current
detection threshold of $R_{\min} = 25~\mu\text{m}$ under standard operating
conditions of the Bruker M6 Jetstream, with the dominant practical limitation
being raster scan time per unit area rather than fundamental contrast
sensitivity. Scaling the methodology to $\sim\!\text{m}^{2}$ exposures appears
within reach of existing instrumentation.

We outlined the geological selection and dating program required to certify a
paleodetector sample, combining $^{87}\text{Rb}/^{86}\text{Sr}$ or U/Pb
geochronology for the primary crystallization age with \textit{in situ}
$^{238}\text{U}$ fission track counting as an internal calibration of the
cumulative thermal annealing history. This dual approach brackets the effective (non-annealed) exposure time
$t_{\text{exp}}$ self-consistently, with the fission-track count in
particular being essential: a sample whose intrinsic fission-track density is
consistent with its isotopic age can be trusted to have retained the larger,
hydrodynamically-driven melt features produced by composite dark matter
transits.

Combining these elements with the Standard Halo Model, we projected
sensitivities for benchmark $1~\text{m}^{2} \times 10^{9}~\text{yr}$ exposures
of muscovite in both the opaque and diffuse composite regimes. We
also revisited the prior monopole-recast exclusions of
Refs.~\cite{Acevedo:2021tbl,Jacobs:2014yca} in the strongly-interacting
regime, and identified a band of parameter space, between the fission-track
retention threshold and the onset of macroscopic melt features, in which
the monopole data can be reliably recast as a dark matter bound, and a complementary band in which
they cannot. The XRF-mapped melt-track and bore-channel search described here
targets part of the latter window using a new etch-free readout approach.

Several extensions of this work are natural. The melt-radius framework should
be extended to composites whose internal structure produces a
velocity-dependent scattering cross section, and to long-range interactions.
The XRF readout demonstration here is a single-step calibration; scan rate,
contrast floor, and angular acceptance as a function of feature size and
sample depth all warrant dedicated further characterization with engineered
sub-micron defect arrays. The treatment of dark matter trajectories through
the Earth at high scattering cross-section neglects gravitational deflection
and possible Earth capture, expected to be relevant only in a limited corner
of the high-mass, high-cross-section parameter space, but should later be
quantified for completeness. Finally, the LMC-induced high-velocity tail
enhancement~\cite{Bozorgnia:2025lmc} and the dynamical history of any
specific mica sample's orientation and burial depth over its retention age
should be folded into any analysis that aims to either translate a future null
result into a robust limit, or to characterize what would be the first
detection of nanoscale composite dark matter: fossilized tracks melted
into gigayear-old mica sheets, waiting to be read out.
\section*{Acknowledgements}
We thank Levente Balogh, Zachary Picker, Tanner Trickle, Aaron Vincent, and Alexis Willson for useful discussions. The authors acknowledge support from the Natural Sciences and Engineering Research Council of Canada (NSERC), the Social Sciences and Humanities Research Council of Canada (SSHRC), the Ontario Early Researcher Award (ERA), and the Canada Foundation for Innovation (CFI). This research was undertaken thanks in part to funding from the Arthur B. McDonald Canadian Astroparticle Physics Research Institute.  Research at Perimeter Institute is supported by the Government of Canada through the Department of Innovation, Science, and Economic Development, and by the Province of Ontario.
\bibliographystyle{JHEP.bst}
\bibliography{biblio.bib}
\end{document}